\documentclass[aps,pra,twocolumn,superscriptaddress,longbibliography]{revtex4-2}

\usepackage{amsmath}
\usepackage{bm}
\usepackage{amsfonts}
\usepackage{mathtools}
\usepackage{graphicx}
\usepackage{braket}
\usepackage{listings}
\usepackage{algorithm}
\usepackage{algorithmic}
\usepackage{color}

\begin{document}

\title{Quantum Zeno approach  for molecular energies with maximum commuting initial Hamiltonians}

\author{Hongye Yu}
\affiliation{Department of Physics and Astronomy, State University of New York at
Stony Brook, Stony Brook, NY 11794-3800, USA}
\author{Tzu-Chieh Wei}
\affiliation{Department of Physics and Astronomy, State University of New York at
Stony Brook, Stony Brook, NY 11794-3800, USA}
\affiliation{C. N. Yang Institute for Theoretical Physics, State University of New York at
Stony Brook, Stony Brook, NY 11794-3840, USA}
\affiliation{Institute for Advanced Computational Science, State University of New York at Stony Brook, Stony Brook, NY 11794-5250, USA}

\begin{abstract}{We propose  to use a quantum  adiabatic and  simulated-annealing framework to compute the ground state of small molecules. The initial Hamiltonian of our algorithms is taken to be the maximum commuting Hamiltonian that consists of a maximal set of commuting terms in the full Hamiltonian of molecules in the Pauli basis. We consider two variants. In the first method, we perform the adiabatic evolution on the obtained time- or path-dependent Hamiltonian with the initial state as the ground state of the maximum commuting Hamiltonian.  However, this method does suffer from the usual problems of adiabatic quantum computation due to  degeneracy and energy-level crossings along the Hamiltonian path. This problem is mitigated by a Zeno method, i.e.,  via a series of eigenstate projections  used in the quantum simulated annealing, with the path-dependent Hamiltonian augmented by a sum of Pauli X terms, whose contribution vanishes at the beginning and the end of the path. In addition to the ground state, the low lying excited states can be obtained using this quantum Zeno approach with equal accuracy to that of the ground state. }
\end{abstract}
\date{\today}
\maketitle

\section{Introduction}
Quantum chemistry concerns the application of quantum mechanics to chemical properties of physical systems, including their electronic structure, spectroscopy, and dynamics~\cite{levine2009quantum,cao2019quantum}. 
It has broad impact, ranging from strongly correlated systems and material design to drug discovery. While mean field theories (such as  density functional theory) have been widely applied to weakly correlated materials, more sophisticated methods, such as the configuration interaction and the density matrix renormalization group, are needed to accurately capture electronic correlation effects~\cite{friesner2005ab,baerends1997quantum,car1985unified,hammond1994monte,chan2011density}. However, the computational cost of these high-level methods scales poorly with the number of orbitals and electrons, thus limiting their applications primarily to small molecules.

Recently it was suggested that quantum chemistry problems could be one of the promising applications for which quantum computation~\cite{nielsen2002quantum} might be used to outperform classical computers. Among the earliest proposals, Lidar and Wang considered using  quantum computation to calculate the thermal rate constant~\cite{lidar1999calculating}. Apsuru-Guzik {\it et al.} proposed to use it 
to calculate the ground-state energy of molecules~\cite{aspuru2005simulated}, which was later implemented experimentally with a photonic system for the hydrogen molecule~\cite{lanyon2010towards}, as well as with a liquid NMR system using an adiabatic state preparation~\cite{du2010nmr}. These and subsequent works prompted a surge of interest in quantum chemistry via quantum computers. The variational quantum eigensolver (VQE) was later proposed for this and implemented in a photonic system~\cite{peruzzo2014variational}. VQE is suitable for noisy intermediate scale quantum (NISQ) processors, as it uses short-depth quantum circuits assisted by classical optimization.  A later study~\cite{o2016scalable} showed that  VQE achieves a better performance than the standard quantum phase estimation (QPE) algorithm~\cite{nielsen2002quantum} 
in the ground-state calculation of molecules. Due to its simplicity and short circuit depth, VQE has since become a standard approach for quantum chemistry applications~\cite{kandala2017hardware}. The extension to  excited states~\cite{colless2018computation,ollitrault2019quantum} has been made and an adaptive approach~\cite{grimsley2019adaptive} has also been  developed.

Apart from circuit-based quantum computation, evolving a quantum system under a suitably designed time-dependent Hamiltonian can also achieve universal quantum computation. This possibility was first proposed by  Kadowaki and Nishimori in the context of quantum annealing to solve the ground-state energy of a classical Ising model~\cite{kadowaki1998quantum} and later by Farhi {\it et al.} using adiabatic evolution to solve instances of the satisfiability problem~\cite{farhi2000quantum}.
It is thus natural to consider adiabatic quantum computation (AQC) for quantum chemistry problems.
Recently, several studies have indeed explored such an application. For example,  adiabatic state preparation was used for studying the hydrogen molecule~\cite{du2010nmr}. In Ref.~\cite{Babbush2014}, a method was introduced to map the Hamiltonians of molecules  to a 2-local qubit Hamiltonian, containing only ZZ, XX and XZ terms.  Another method was recently proposed  in Ref.~\cite{xia2017electronic} to map  quantum-chemistry Hamiltonians to Ising spin-class Hamiltonians, and   experiments based on this method were  carried out on the existing quantum annealers of D-Wave~\cite{streif2019solving}. 

Here, we present a study of molecular energies via an adiabatic framework that can be used  in two different ways.  Both variants are based on a path-dependent Hamiltonian as used in  adiabatic quantum computation~\cite{farhi2000quantum}. The initial Hamiltonian is constructed using a maximal set of the commuting Pauli terms (in the qubit version) in the molecular Hamiltonian as described in details below. We will refer to this as the maximum commuting (MC) Hamiltonian. (It turns out that every term in the MC Hamiltonian is proportional to a product of Pauli Z and identity operators.) We note that similar ideas of such grouping via commuting terms also appeared in previous works to reduce the cost of  measurements~\cite{1907.13623,doi:10.1021/acs.jctc.0c00008,verteletskyi2020measurement,jena2019pauli,izmaylov2019unitary}, 
as well as in Ref.~\cite{kirby2020classical} in the context of  `noncontextual' Hamiltonians and their classical simulations. The time-dependent Hamiltonian for our purpose is taken to be the linear interpolation (for time $t\in[0,T]$)  between the maximum commuting Hamiltonian and the molecular Hamiltonian, i.e., with weights $(1-t/T)$ and $t/T$, respectively. 
 
 Our first variant  is the usual adiabatic quantum computation with such a time-dependent Hamiltonian.  
This approach of quantum adiabatic  evolution (QAE)  yields accurate results of molecular ground-state energies around the equilibrium atomic position or  with small perturbations of the bond length, but not at the limit of bond breaking, which is due to the degeneracy of ground and excited states, as well as energy-level crossings. These issues arise because at large  inter-atomic separations, there are many closely spaced energy levels. In an attempt to ameliorate these issues in obtaining the ground state and its energy, we add to the Hamiltonian a possible degeneracy breaking term whose strength is proportional to $(1-t/T)t/T$, so that its contribution vanishes at the initial and final times~\cite{farhi2002quantum}.
Even though the resultant path-dependent Hamiltonian may not necessarily yield better results under adiabatic evolution, its use in the setting of the quantum simulated annealing~\cite{somma2008quantum} does improve the obtained ground-state energy. The latter is thus our second variant, which drives the computation via the  quantum Zeno-like projection (QZP) to eigenstates of the instantaneous Hamiltonian along the path at discrete  time steps~\cite{somma2008quantum,poulin2009preparing,chen2020quantum}. This QZP method, with the augmented time-dependent Hamiltonian,  can mitigate the drawbacks  of  adiabatic evolution.   By starting  the initial state to be the ground state or lowest few excited states of the maximum commuting Hamiltonian, the ground state and the lowest few excited states of the final Hamiltonian can be obtained for several small molecules that we consider, including LiH,  Be$\text{H}_2$, CH$_2$, and $\text{H}_2$O. We also compare our results with those from the VQE. Our numerical simulations show that the QZP method performs the best among the three methods.
 
The paper is organized as follows. In Sec.~\ref{sec:MaxGreedy} we introduce the concept of the maximum commuting Hamiltonian and give a greedy algorithm that can efficiently  approximate the maximal commuting set. In Sec.~\ref{sec:AQC} we use  adiabatic quantum computation  to drive the system to the molecular ground state  and numerically show that it works well in most cases and regimes of interest. In Sec.~\ref{sec:Projection} we propose to use a spectral projection method for improving the results of the QAE with the time-dependent Hamiltonian augmented by a Pauli X term. This also allows us to obtain excited states without further complexity. In Sec.~\ref{sec:HF}, we discuss an alternative way to construct the time-dependent Hamiltonian,  using the Hartree-Fock Hamiltonian as the initial Hamiltonian. But we argue that this procedure is not as favorable as the maximum commuting Hamiltonian. In Sec.~\ref{sec:conclude} we make concluding remarks.

\section{Maximum Commuting Set and Greedy Approximation}
\label{sec:MaxGreedy}
In this section, we first discuss the molecular Hamiltonian and the associated atomic orbitals, and then introduce the maximum commuting Hamiltonian and its obtainment from the qubit-version of the molecular Hamiltonian via a greedy algorithm.
\subsection{Molecular Hamiltonian}
The coefficients in the Hamiltonians for the molecules, LiH, Be$\text{H}_2$, $\text{H}_2$O, and CH$_2$ that we consider in this work are computed in the Slater-type orbital (STO)-3G basis (see e.g.~\cite{levine2009quantum}). For the LiH molecule, we shall measure the distance as the inter-atomic distance between Li and H atoms, and arbitrarily assign this direction as the $x$ axis. The orbitals that will be used in the calculation include  $1s$ for the H atom and $1s, 2s$, and $2p_x$ for the Li atom.
For the tri-atomic molecules of the form $\text{AB}_{2}$, including Be$\text{H}_2$, CH$_2$ and $\text{H}_2$O, we will assume equal distance of the two B atoms measured from the A atom and study the molecular energy as a function of the distance $\overline{\text{AB}}$ at a fixed $\angle$BAB angle or as a function of the $\angle$BAB angle at a fixed distance, typically at the equilibrium position.  The orbitals that will be used in the calculation of the tri-atomic molecules include  $1s$ for the H atom and  $1s, 2s, 2p_x, 2p_y$, and $2p_z$ for Be, C and, O atoms. We assume that for the LiH molecule the $2p_y$ and $2p_z$ orbitals can be ignored due to their linear configuration,  but these are included in our calculations for Be$\text{H}_2$, CH$_2$, and $\text{H}_2$O molecules.  Our goal is to find the eigenstates and  eigenenergies of a molecule's Hamiltonian $H$, especially the ground state $\ket{\psi_G}$ and its energy $E_G$,   with
\begin{equation}
H\ket{\psi_G}=E_G\ket{\psi_G},
\end{equation}
as well as a few low lying states and associated energies. 

In the spin-orbital language, the Hamiltonian can be written in  second quantized  form as
\begin{equation}
H=H_{1}+H_{2}=\sum_{\alpha, \beta} t_{\alpha \beta} a_{\alpha}^{\dagger} a_{\beta}+\frac{1}{2} \sum_{\alpha, \beta, \gamma, \delta} u_{\alpha \beta \gamma \delta} a_{\alpha}^{\dagger} a_{\gamma}^{\dagger} a_{\delta} a_{\beta},
\end{equation}
where the Greek subscripts label the orbitals and the coefficients in the one-body and two-body terms are given below, respectively,
\begin{equation}
\label{eq:spin-orbital}
\begin{aligned} t_{\alpha \beta} &=\int d x_{1} \Psi_{\alpha}\left(x_{1}\right)\left(-\frac{\nabla_{1}^{2}}{2}+\sum_{i} \frac{Z_{i}}{\left|r_{1 i}\right|}\right) \Psi_{\beta}\left(x_{1}\right), \\ u_{\alpha \beta \gamma \delta} &=\iint d x_{1} d x_{2} \Psi_{\alpha}^{*}\left(x_{1}\right) \Psi_{\beta}\left(x_{1}\right) \frac{1}{\left|r_{12}\right|} \Psi_{\gamma}^{*}\left(x_{2}\right) \Psi_{\delta}\left(x_{2}\right). \end{aligned}
\end{equation}
These  are calculated using a standard quantum chemistry package, such as {\tt PySCF}~\cite{PYSCF},  which is a  collection of electronic structure programs powered by {\tt Python}.
In applying Eq.~(\ref{eq:spin-orbital}), we will assume that all $1s$ orbitals of Li, Be, C, and O atoms are filled and inert, which means that some of the one-body integrals will become constant and only brings a shift to the total energy and  some of two-body integrals will reduce to one-body terms.
\subsection{Maximum Commuting Hamiltonian}
In order for our algorithms to be performed on   quantum computers, we need to convert the above fermionic Hamiltonian to its qubit version. There have been a few ways proposed to do this, including  Jordan-Wigner, parity, Bravyi-Kitaev, and superfast Bravyi-Kitaev transformations~\cite{seeley2012bravyi,setia2018bravyi,setia2019superfast}. By using any of these methods, we can transform the fermion operators into Pauli operators,
\begin{equation}
\label{eq:Hp}
	H_p=\sum_i h_i P_i,
\end{equation}
where $P_i$'s are $n$-qubit Pauli operators, which can contain qubit identity operators in the product. For the conversion below, we use the `parity' method.

Next, we focus on the adiabatic-based framework and first introduce a special initial Hamiltonian called the maximum commuting (MC) Hamiltonian. It will be demonstrated below that the ground state of the molecular Hamiltonian can be found in the adiabatic approach by connecting this initial MC Hamiltonian to the final Hamiltonian of the molecule. This procedure works well at least for distances around the equilibrium position. Below we  will also discuss the issues that this approach will encounter, and later provide an alternative method to ameliorate them.

\begin{figure}[t]
	\centering
	\includegraphics[width=0.5\textwidth]{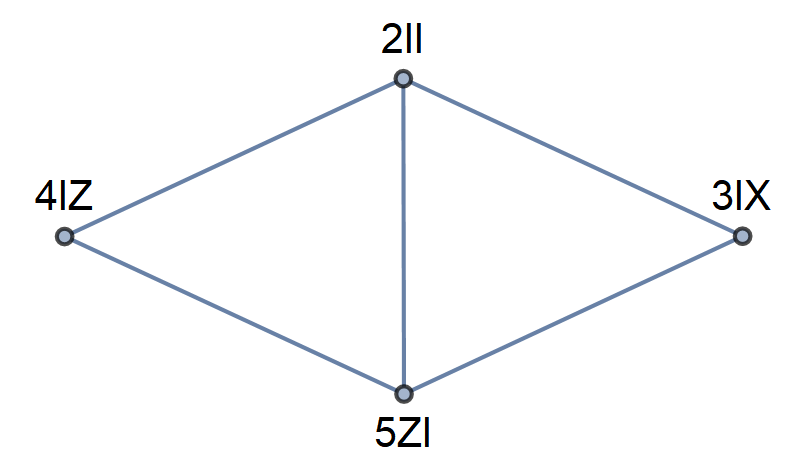}
	\caption{The weighted graph constructed from $H=2II+3IX-4IZ+5ZI$. Two vertices representing operators $P_i$ and $P_j$ are connected by an edge if $[P_i,P_j]=0$.}
	\label{graph}
\end{figure}

We use the symbol $\mathcal{S}$ to denote a set of  Pauli product operators that commute with  each other,
\begin{equation}
\mathcal{S}\equiv\{P_1,P_2,...,P_k~|~\forall i,j \in \{1,...,k\},[P_i,P_j]=0\},
\end{equation}
where $P_j$'s come from terms in the qubit molecular Hamiltonian~(\ref{eq:Hp})~\cite{nielsen2002quantum}.
A maximum commuting Hamiltonian associated with a qubit-based Hamiltonian $H=\sum_i h_i P_i$ is a maximal commuting set $\mathcal{S}^*$ that maximizes the weight 
\begin{equation}
\max_{\mathcal{}{S}}\sum_{i;\,P_i\in \mathcal{S}} |h_i|
\end{equation}
among all possible commuting sets. Here we give a simple example to illustrate the  maximal commuting set. Consider the following Hamiltonian,
\begin{equation}
H=2II+3IX-4IZ+5ZI,
\end{equation}
we assign every Pauli operator $P_i$ in $H$ to a vertex, and associate with it a weight $|h_i|$ and every pair of two commuting Pauli operators to an edge. We note the un-weighted version of the graph was used in the context of contextuality; see, e.g., Ref.~\cite{kirby2019contextuality}. The Hamiltonian is  thus represented by a weighted graph, as illustrated in Fig.~\ref{graph}.
The maximal commuting set of $H$ is the maximum weighted `clique' of the corresponding graph, which consists of $\mathcal{S}^*=\{II,IZ,ZI\}$ and has a weight of $11$ in this example. The set $\{II,IX,ZI\}$, on the other hand, has a smaller weight of $10$.

\begin{figure*}[t!]
	\centering
	\includegraphics[width=0.9\textwidth]{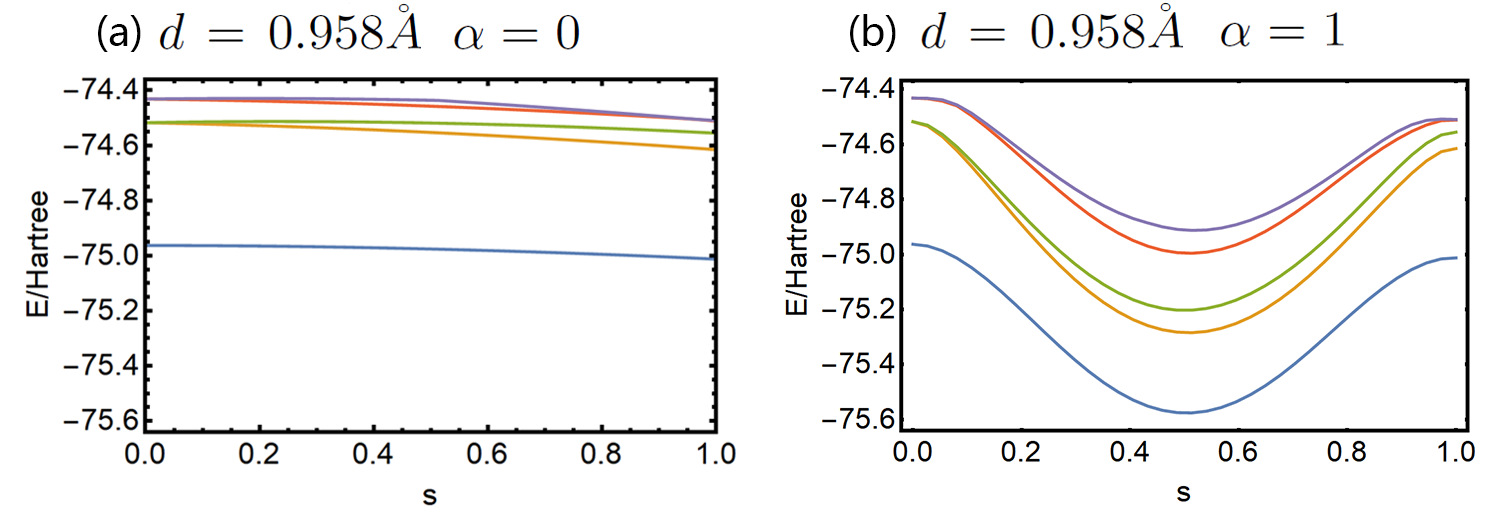}
	\caption{The lowest few energy levels of $H_\alpha(s=t/T)$ of Eq.~(\ref{eq:Halpha}) for  $\text{H}_2$O  at the equilibrium position $d=0.958 \AA{}$ between the O and an H atom with  (a) $\alpha=0$ and (b) $\alpha=1$. In case (a), the Hamiltonian reduces to $H(s=t/T)=(1-s)H_i+s\,H_p$ of Eq.~(\ref{eq:H}). }
	\label{gap}
\end{figure*}
\begin{figure*}[t!]
	\centering
	\includegraphics[width=0.99\textwidth]{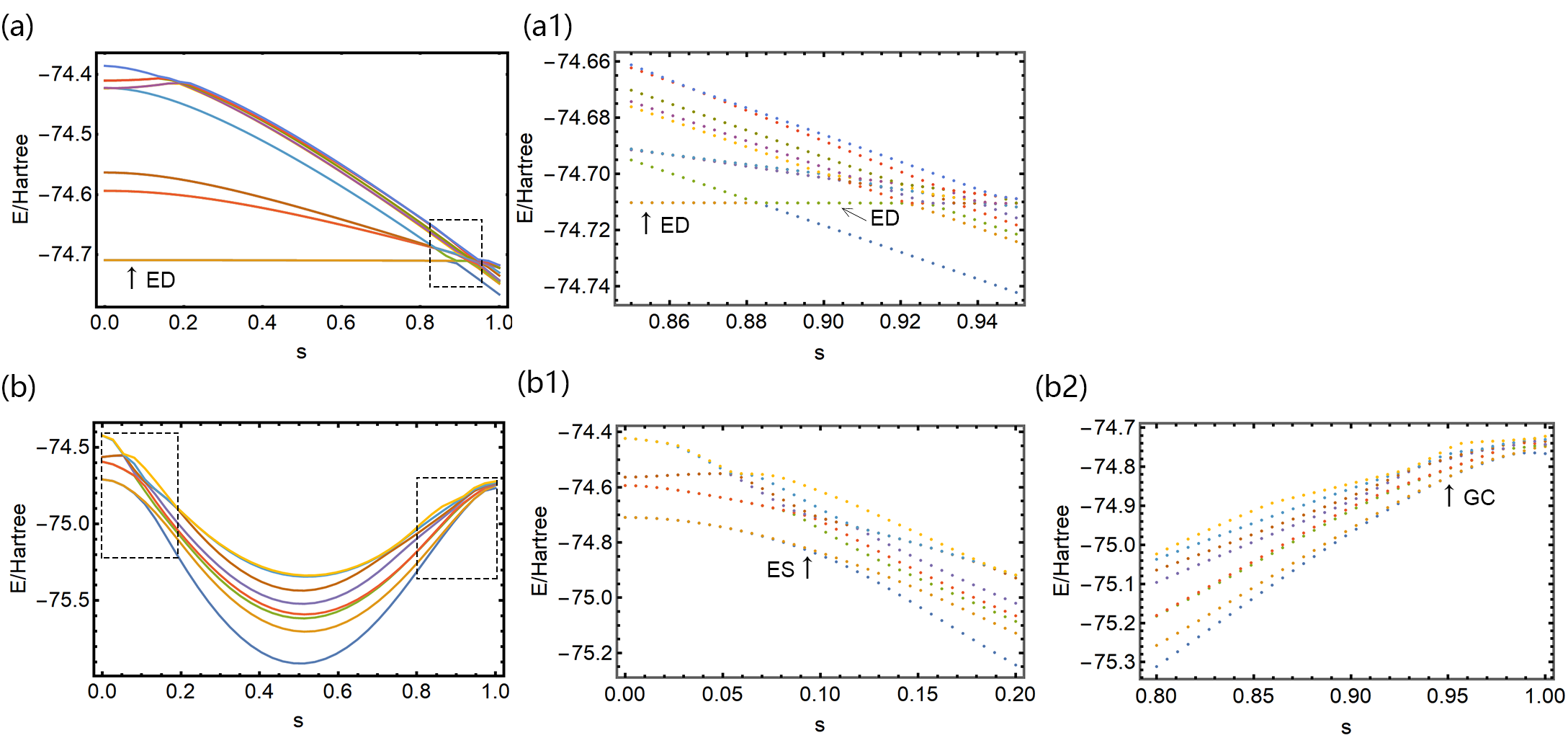}
	\caption{The lowest few energy levels of $H_\alpha(s=t/T)$ of Eq.~(\ref{eq:Halpha}) for  $\text{H}_2$O at a position $d=1.958 \AA{}$ with (a) $\alpha=0$ and (b) $\alpha=1$. Notice that the ranges of the values in the vertical axes are different. Figures (a1) and (b1) \& (b2) show the blow-up in \textcolor{black}{the  boxed} regions of (a) and (b), respectively, and the {\color{black} data are represented by dots instead of connected curves}.   In (a), the initial ground states have a two-fold energy degeneracy (ED) and are crossed by another energy level near $s=0.885$. 
	After adding the $H_X$ term, see Eq.~(\ref{eq:Halpha}), the doubly degenerate levels gradually split, as indicated by ES in (b1). As seen in (b2), the splitting reaches a maximum around $s=0.5$ and then gradually closes, as indicated by gap closing (GC). The gap closes at around $s=0.95$  but it slightly opens up towards $s=1$. 
	}
	\label{gap1}
\end{figure*}
\subsection{Greedy  approximation algorithm}
\label{subsec:greedy}
It is in general not an easy task to find the maximal commuting set, as it is equivalent to the weighted maximum clique problem. While the problem is NP-hard~\cite{bomze1999maximum}, a greedy algorithm can give a good approximation efficiently when weights are highly biased. For the purpose of molecular energies, we do not require the set of commuting Pauli terms to be the exact maximum. Our greedy algorithm is given below.

\begin{algorithm}[H]
	\caption{Greedy algorithm for maximal clique}
	\begin{algorithmic}
		\STATE \textbf{input}: a weighted graph $G$
		\STATE \textbf{output}: a maximal clique of $G$
		\STATE \textbf{begin}
		\STATE set $S=\emptyset, I=V(G)$, where $V(G)$ is the vertex set of $G$
		\WHILE{$I$ is not empty}
		\STATE find the vertex in $I$ that has a maximum weight, say $v_i$
		\STATE $S=S\bigcup \{v_i\}$
		\STATE $I=I\setminus\{v_i\}\setminus V_{s_i}$, delete $v_i$ and all vertices that are not connected to every vertex in $S$, say $V_{s_i}$, in $I$
		\ENDWHILE
		\RETURN $S$
		\STATE \textbf{end}
	\end{algorithmic}
\end{algorithm}

After performing this greedy algorithm, we obtain one maximal clique of the graph, which is in general not necessarily the absolute maximum clique. But for the four molecules (LiH, $\text{H}_2$O, Be$\text{H}_2$, and CH$_2$) that we shall simulate, the maximal cliques found by our greedy algorithm turn out all to be the maximal ones. Even if we do not obtain the absolute maximum clique, the one generated by the greedy algorithm can still be used as an initial Hamiltonian. Note that the above greedy algorithm has only linear-time complexity. So in practice it is efficient in finding a good approximate maximal clique for the problem of simulating molecular energies.

Once the initial MC Hamiltonian is obtained, we need to be able to initialize the system in its ground state (or other eigenstates) for  adiabatic quantum computation. 
We note that, however, commuting Hamiltonians are not necessarily easy to solve. Classical Ising spin glass models are such an example and are NP-complete, as shown by Barahona~\cite{barahona1982computational}. Bravyi and Vyalyi showed that generic quantum-mechanical 2-local commuting Hamiltonian problems are  also NP-complete~\cite{bravyi2005commutative} and Aharonov and Eldar did the same for 3-local commuting Hamiltonian problems~\cite{aharonov2011complexity}. However, there are commuting Hamiltonians that are easy. For example, models such as Kitaev's toric code~\cite{kitaev2003fault} and Levin and Wen's string-nets~\cite{levin2005string} are solvable. Furthermore, Yan and Bacon showed that $k$-local Hamiltonians with projectors
onto eigenspaces of product of Pauli matrices are in the complexity class P~\cite{yan2012k}.  
In our consideration of molecular Hamiltonians, we find that those terms in the MC set are all of the  form of a product of Pauli Z and identity operators with a certain weight, such as $ c \,I\otimes \sigma^z \otimes \sigma^z\otimes\cdots\otimes I$.  We note that the complexity consideration regarding the NP-completeness deals with the worst-case scenario. Very often the claim of such hardness requires inclusion of certain randomness, such as a random local field or some random couplings~\cite{schuch2009computational}. We cannot precisely characterize the complexity of our MC Hamiltonian, despite the fact that  the generic spin glass problem is NP-complete. As a remark, it is believed by complexity theorists that even NP-complete problems cannot be efficiently solved even by quantum computers~\cite{aaronson2008limits}.

 Nevertheless, given that our MC Hamiltonian is diagonal in the computational basis, it is essentially a classical problem. By substituting the Pauli-Z operators by Boolean variables $\sigma^z_i=2x_i-1$, with $x_i=0$ or $1$, we can transform  our MC Hamiltonian to a Boolean optimization problem. This is a weighted Max-2SAT form if the initial reduction from fermions to spins is done with Jordan-Wigner transformation (instead of the parity mapping). While generic Max-2SAT problems are still NP-complete, there exist some efficient classical algorithms such as MiniMaxSAT~\cite{heras2008minimaxsat} and wMaxSATz~\cite{li2009exploiting} for average cases. Although the rigorous time complexity of these algorithms is unknown and their performance varies from case to case, the time complexity is at most $o(N=2^{n})$, where $n$ is the total number of qubits. (Given the small number of qubits in the problems we consider,  we can directly solve the Boolean optimization or the
original commuting Pauli Z problems.) However, we note that the time complexity of finding the ground state of the final full Hamiltonian (of dimension $N\times N$) is of order $O(N^3=2^{3n})$. For a sparse matrix, this can be reduced to $O(s N^2)$, where $s$ is the average number of nonzero elements in a row or column.  We can estimate $s$ by the number of terms in the Hamiltonian and it is of order $n^4$. Therefore, the time complexity for solving the ground state of the final quantum chemistry problem scales at least $O(n^42^{2 n})$. Thus,  there is still speedup in solving it using our proposed quantum algorithms below, compared to classical means.

 Moreover, our first approach (see Sec.~\ref{sec:AQC}) relies on the exact ground state of the MC Hamiltonian, whereas the second approach (see Sec.~\ref{sec:Projection}) can work with an approximate ground state that  has non-unity but finite nonzero overlap with the exact MC Hamiltonian ground state.  Thus, the latter approach may have better time complexity and speedup compared with classical algorithms.  
 
\section{The approach by quantum adiabatic evolution}

\label{sec:AQC}
In this section we introduce our first quantum algorithm for the  molecular ground-state energy, by adiabatically evolving a time-dependent Hamiltonian, listed in Eq.~(\ref{eq:H}). We then present the results obtained from applying the algorithm to several molecules and discuss some issues this algorithm encounters.
\subsection{Adiabatic evolution}

To drive the system from the ground state of the maximum commuting Hamiltonian to that of the desired Hamiltonian, a natural way is to use the idea of adiabatic evolution~\cite{farhi2000quantum}. From the maximal commuting set $\mathcal{S}^*$ associated with the molecular Hamiltonian~(\ref{eq:Hp}), we have the corresponding maximum commuting Hamiltonian  
\begin{equation}
\label{eq:MSH}
H_i=\sum_{i; P_i\in \mathcal{S}^*} h_i P_i.
\end{equation}
We take $H_i$ to be the initial Hamiltonian and   $H_p$ the final Hamiltonian, and form 
the following  time-dependent Hamiltonian
\begin{eqnarray}
H(t)&=&(1-\frac{t}{T})H_i+\frac{t}{T}H_p\nonumber\\
&=&\sum_{i;P_i\in \mathcal{S}^*} h_i P_i+\frac{t}{T}\sum_{i;P_i\notin \mathcal{S}^*} h_i P_i.\label{eq:H}
\end{eqnarray}

\begin{figure}[t]
	\centering
	\flushleft (a)\\ \includegraphics[width=0.5\textwidth]{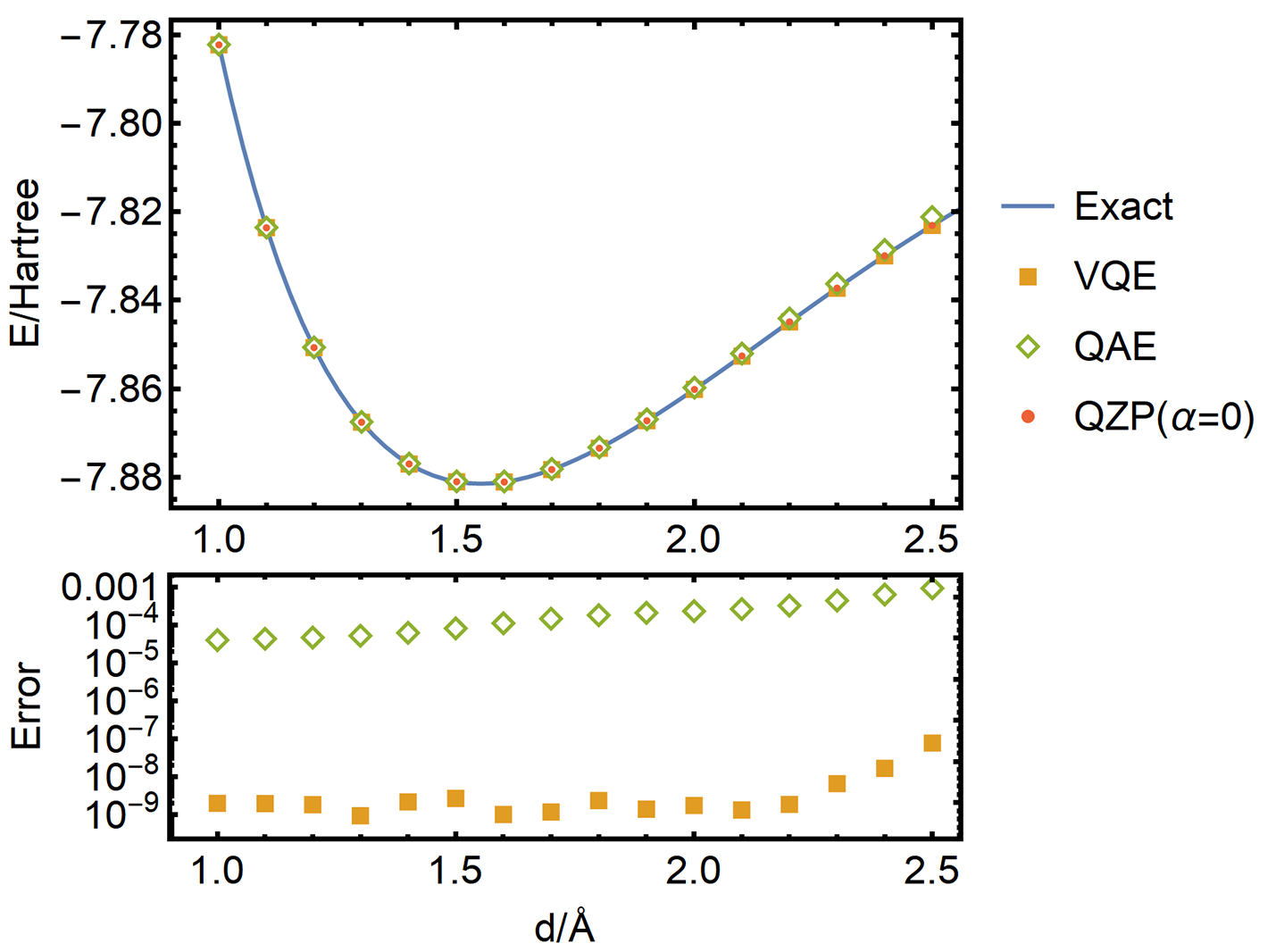}
	(b)	\\ \includegraphics[width=0.5\textwidth]{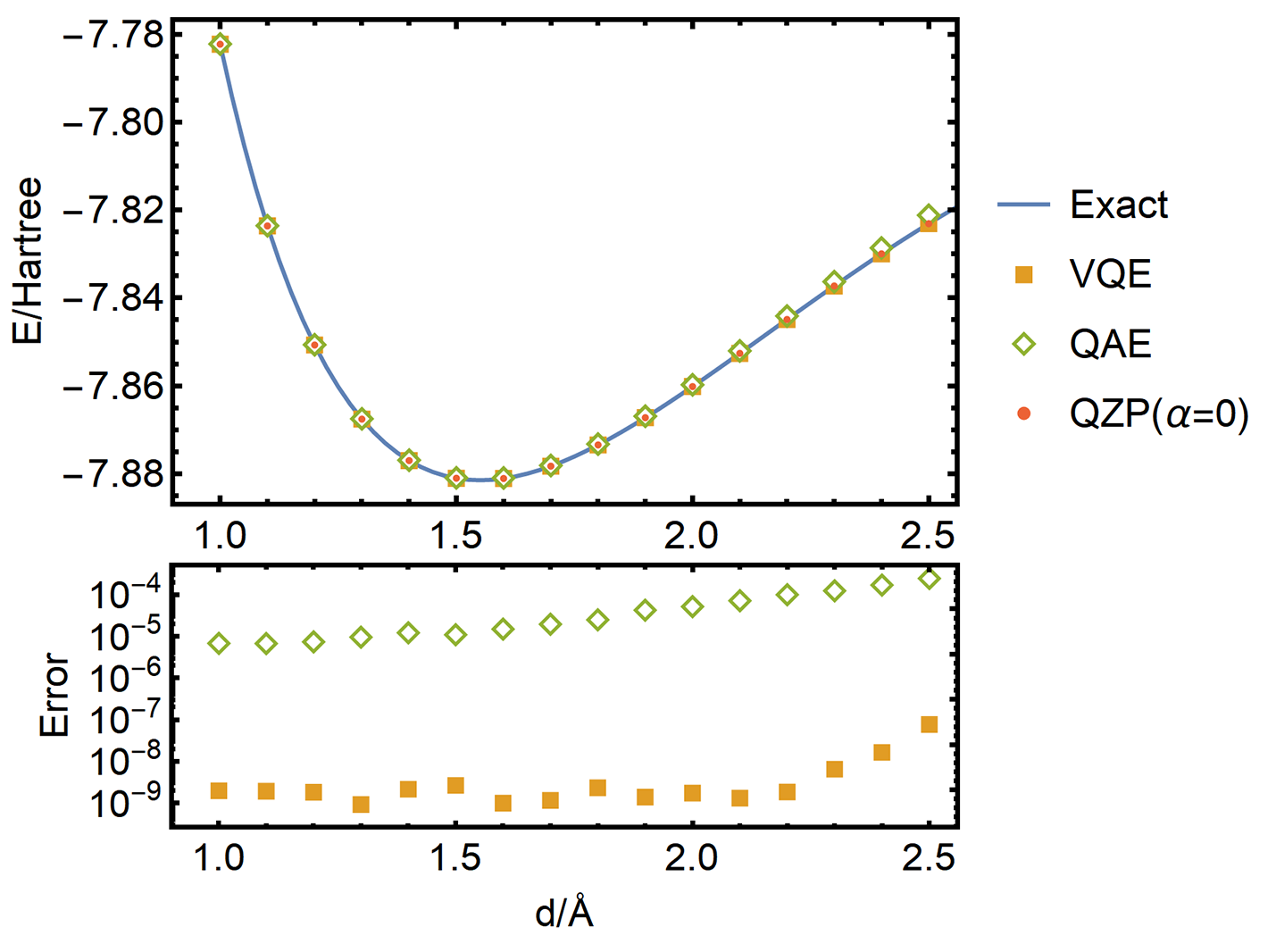}
	\caption{The ground state energy of LiH with different approaches---(a) top panel: results using $T=10$ and $\Delta T=0.5$, (b) bottom panel: results using $T=40$ and $\Delta T=0.5$. The result of QAE is slightly worse than that of the VQE because of the small size of the 4-qubit Hamiltonian, but can be improved using longer time or equivalently more segments. Note the QZP points are not shown in the error figure as they are 0-error to the machine precision.}
	\label{LiH}
\end{figure}
We  set the initial state $\ket{\psi(0)}$ to be the ground state of $H_i$ and let the system evolve under the time-dependent Hamiltonian $H(t)$. According to the adiabatic theorem~\cite{teufel2001note}, if the system evolves slowly enough, i.e., $T$ being sufficiently large, then the evolved state $\ket{\psi(t)}$ will stay very close to the instantaneous ground state of $H(t)$, provided that there is a finite gap separating the ground state from the excited states. Roughly speaking, $T$ should scale inversely proportional to the square of the minimal gap along the path, i.e., $\sim 1/\Delta^2$. At the end of the evolution, the system will arrive at our desired state---the ground state of $H_p$, up to a small error. We call this procedure the quantum adiabatic evolution (QAE). 
%If the first term $\sum_{i;P_i\in \mathcal{S}} h_i P_i$ of $H(t)$ in Eq.~(\ref{eq:H}) includes large coefficients, which is true in most cases, the second term may be regarded as a perturbation. 

In Fig.~\ref{gap}a, we illustrate the lowest few energy levels of the time-dependent Hamiltonian $H(t)$  as a function of $s=t/T$ for the water molecule at its equilibrium position.  The gap between  the ground state and first excited state
remains roughly constant along the path, which is the most favorable scenario for AQC.
%of the path-dependent Hamiltonian determines the evolution time $T$ (which is roughly $1/\Delta^2$) in order to achieve adiabaticity.
We note that, however, there can be energy level crossings, e.g. in Fig.~\ref{gap1}a for the $\overline{\text{OH}}$ distance roughly twice as large as in Fig.~\ref{gap}a. Then the evolution cannot guarantee  the system  to stay in its instantaneous ground state. But this seems to occur only at large molecular distances; see also discussions in Sec.~\ref{sec:degeneracy}.

\begin{figure}[t]
	\centering
	\includegraphics[width=0.5\textwidth]{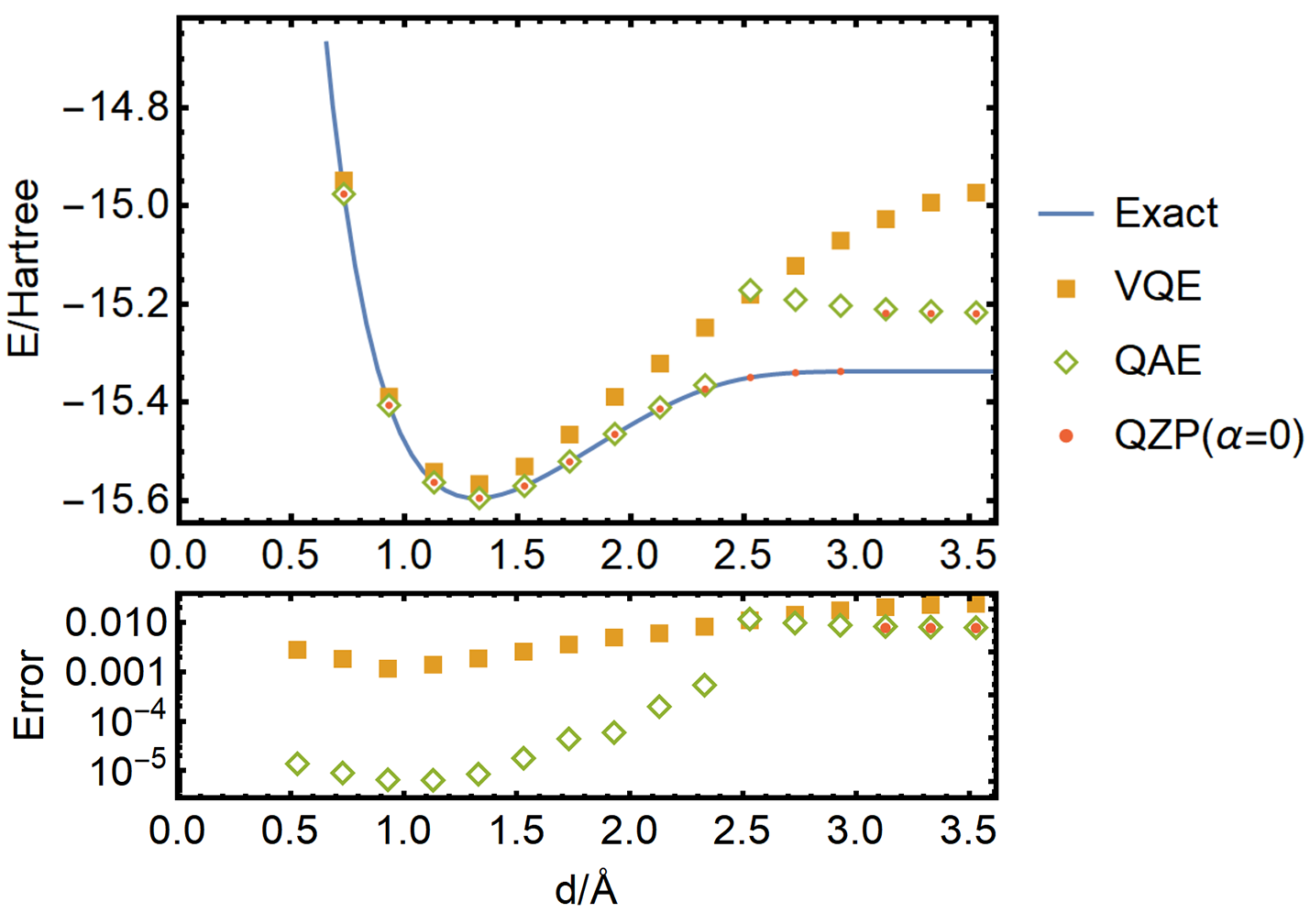}
	\caption{The ground state energy of Be$\text{H}_2$ vs. the distance $d$ between the Be and one H atoms, with different approaches at the equilibrium angle $180^\circ$. The results of the last 6 points of QAE are not as good as the other points due to the appearance of degeneracies in the initial ground states. The lower panel indicates the errors deviating from the exact calculation. Some of the  points  by QZP not shown in the error figure are 0-error points.
}
	\label{BeH2}
\end{figure}
\begin{figure}[t]
	\centering
	\includegraphics[width=0.5\textwidth]{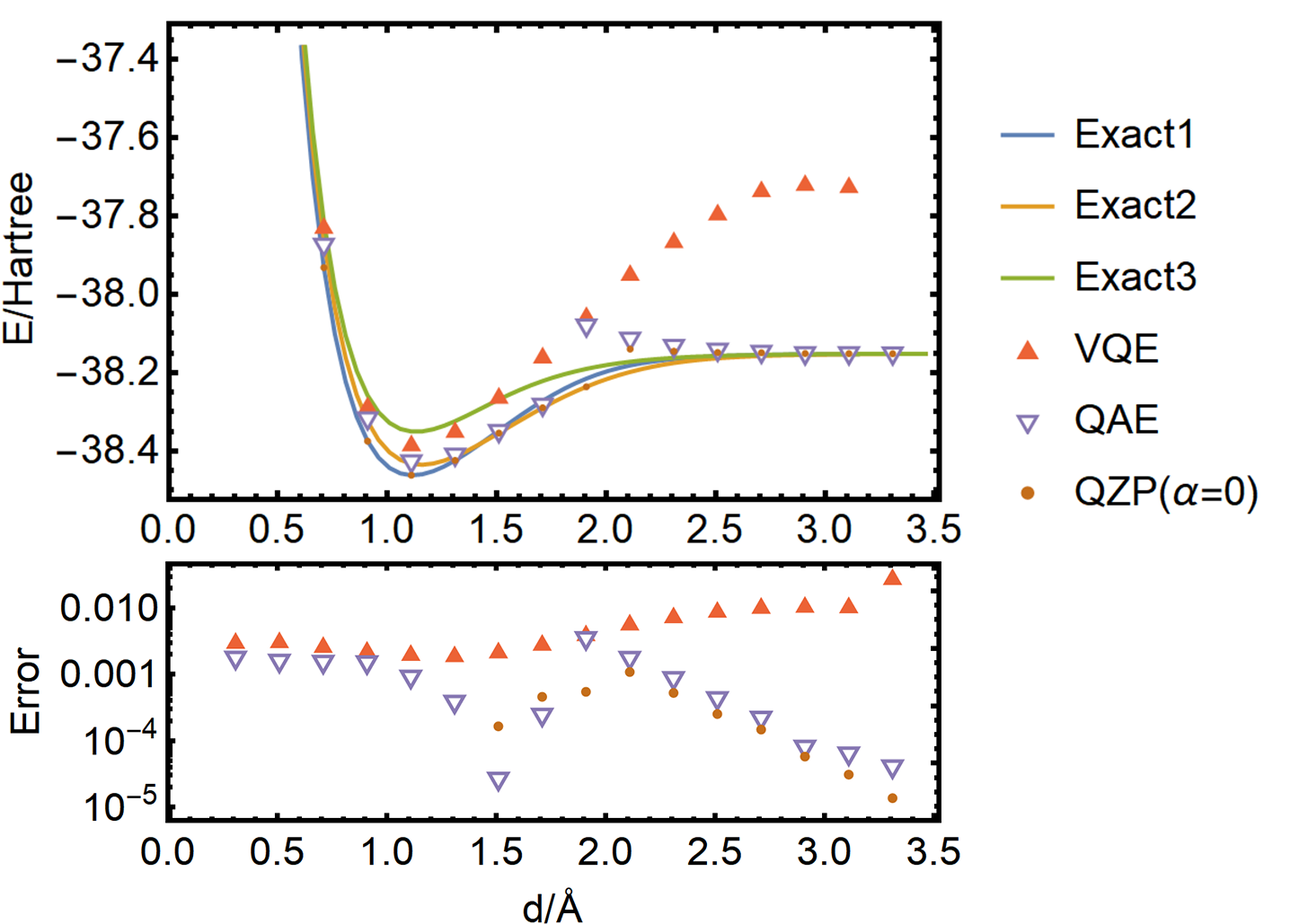}
	\caption{The ground state energy of C$\text{H}_2$ as a function of the distance $d$ between the C and an H atom at the equilibrium angle $\theta\approx101.89^\circ$.  The three curves (labeled 
	`Exact') represent the lowest three energy levels calculated from exact digaonalization, and two of them cross at a distance of around $1.41\AA$.  Some of the QZP points  not shown in the error figure are 0-error points.}
	\label{CH2}
	\end{figure}

\begin{figure}[t]
	\flushleft (a) 
	\includegraphics[width=0.5\textwidth]{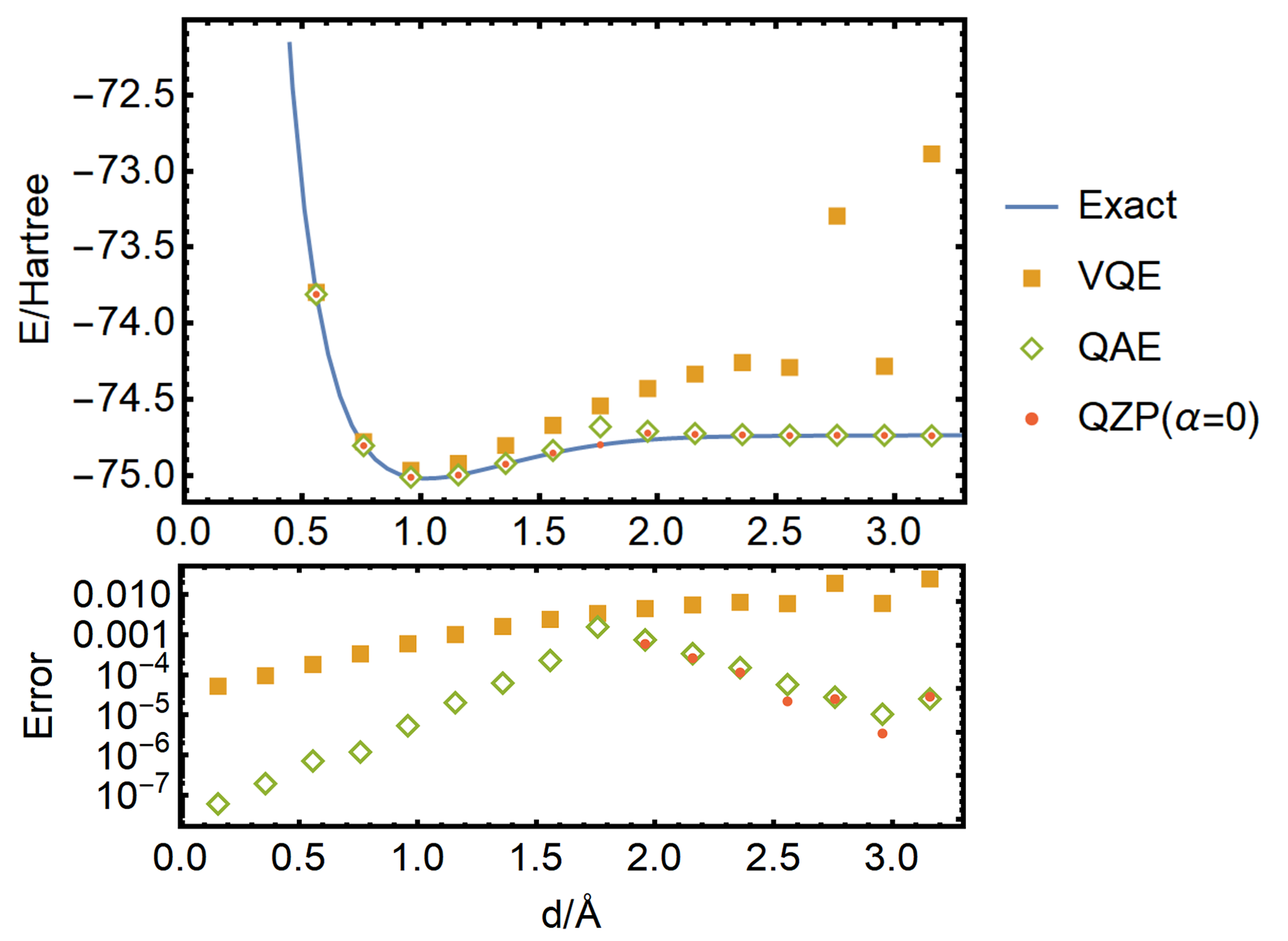}\\
	(b) \includegraphics[width=0.5\textwidth]{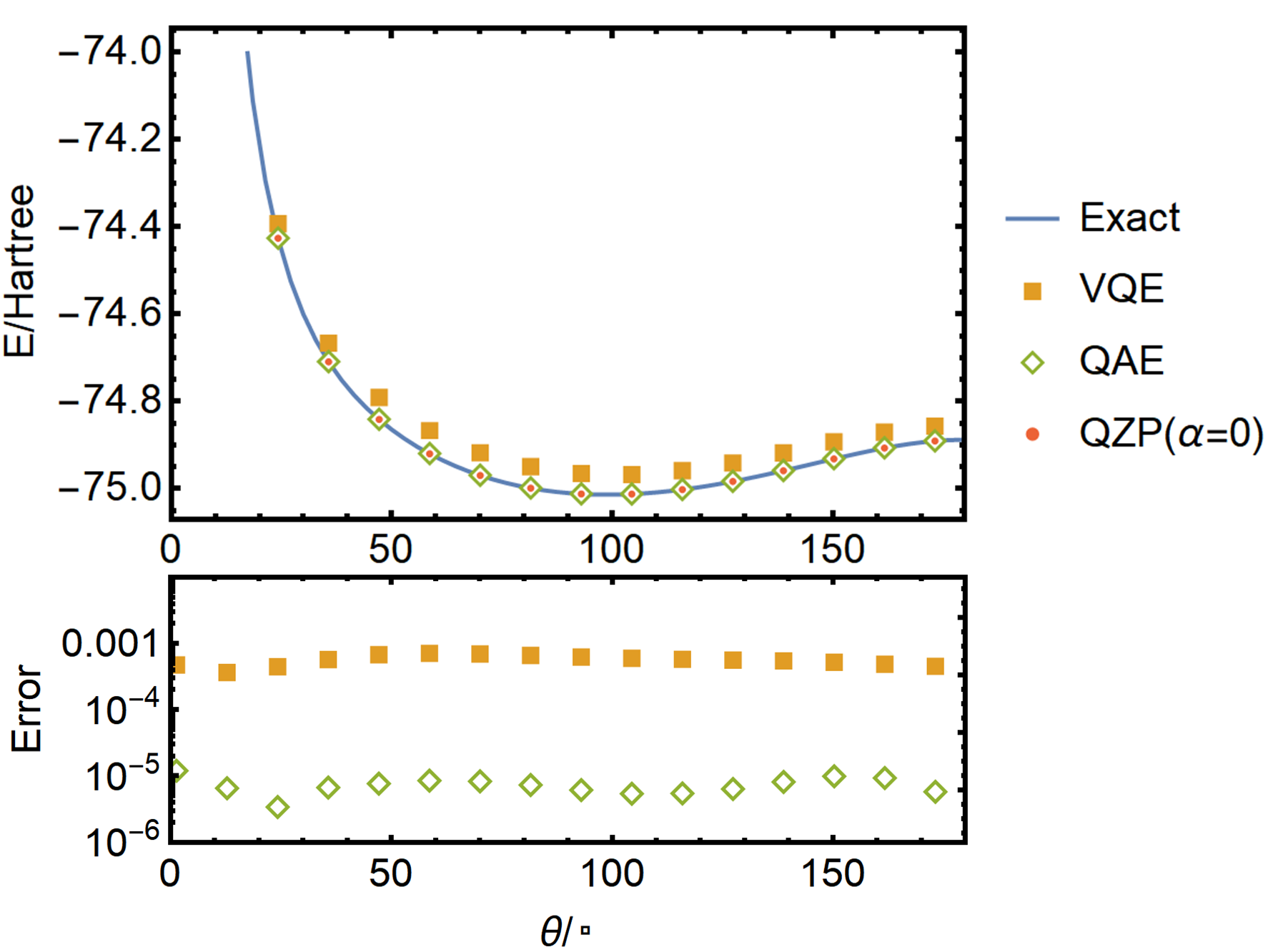}
	\caption{The ground state energy of $\text{H}_2$O (a) as a function of the distance $d$ between the O and an H atom at the equilibrium angle $\theta\approx104.45^\circ$ and (b) a function of the H-O-H angle $\theta$ at the equilibrium position $d=0.958 \AA{}$. The result of QAE at $d\approx 1.76\AA$ is not particularly good because of the  appearance of degeneracies in the initial ground states. The lower panel in both (a) and (b) indicates the errors to the exact calculation. Some of the QZP points not shown in the error figure are 0-error points.}
	\label{H2O}
\end{figure}
\subsection{Ground state energy results}
We apply our QAE approach to four different molecules $\mathrm{LiH},\mathrm{Be}\mathrm{H}_2$, CH$_2$, and $\mathrm{H}_2\mathrm{O}$, and present our results in Figs.~\ref{LiH},~\ref{BeH2},~\ref{CH2}, and~\ref{H2O}, respectively. In  these simulations we set $T=10$  and use discrete-time slices to approximate the continuous time evolution,
\begin{equation}
\label{eq:Adiabatic}
	\ket{\psi(T)}\approx e^{-i H(T)\Delta T}e^{-i H(T-\Delta T)\Delta T}\dots e^{-i H(\Delta T)\Delta T}\ket{\psi(0)}.
\end{equation}
We choose a constant increment $\Delta T=0.5$ in the simulations, as we do not and cannot rely on the knowledge of the gap, which may even close (e.g. see Fig.~\ref{gap1}a).  

Our results from this QAE method are compared to those from  VQE. 
The latter uses certain kinds of ansatz wave function via quantum circuits and performs measurements to read out the energy expectation value. A classical optimizer is then used to search for optimal parameters in the ansatz that minimize the energy expectation value~\cite{peruzzo2014variational,kandala2017hardware}. The quality of VQE depends on the variational ansatz used and the convergence depends  on the classical optimizer. In our VQE calculation we use the UCCSD variational form and the COBYLA optimizer, implemented with the IBM {\tt Qiskit} package~\cite{Qiskit}. It begins with the Hartree-Fock state and runs with iteration steps up to 200 for short molecular distances and 1000 for large molecular distances in order to optimize the variational parameters of the quantum circuit. As seen in Figs.~\ref{BeH2},~\ref{CH2}, and~\ref{H2O}, the VQE results deviate from the exact solutions substantially at large molecular distances, and the computation at certain distances do not even converge properly. The exact solutions we compare to are obtained from directly diagonalizing the associated qubit Hamiltonian in Eq.~(\ref{eq:Hp}).

Below we discuss the results from our QAE for individual molecules.
\subsubsection{LiH molecule}
For LiH molecule, the results of the adiabatic approach are slightly worse than those of the VQE, as seen in Fig.~\ref{LiH}a. This is mostly due to the small number of the discrete-time steps, and the accuracy can be improved by using smaller $\Delta T$ or larger $T$ and hence more  segments along the Hamiltonian path, as demonstrated in Fig.~\ref{LiH}b.  In this calculation, four qubits are needed to represent the Hamiltonian. We note that the simulation of an iterative quantum phase estimation for the ground-state energy of LiH was done in Ref.~\cite{aspuru2005simulated} and the implementation of the VQE on quantum computers was presented in Ref.~\cite{kandala2017hardware}.

\subsubsection{$\text{Be}\text{H}_2$ molecule}
For Be$\text{H}_2$, the results shown in Fig.~\ref{BeH2} are mostly better than those of the VQE, and agree very well with the exact solutions at smaller distances and around the equilibrium position, but become worse at large distances, in particular greater than $d\gtrsim 2.4\AA$. This is due to the degeneracy of ground states in the initial Hamiltonian. (We note that this issue can be ameliorated by employing the projection method with an augmented Hamiltonian, as discussed below.) In this calculation, 10 qubits are needed to represent the Hamiltonian. If we remove the $2p_y$ and $2p_z$ orbitals, as in the case of linear configurations of the three atoms H-Be-H, then the number of qubits can be reduced to 6. We remark that the implementation of the six-qubit VQE on quantum computers was previously presented in Ref.~\cite{kandala2017hardware}.

\subsubsection{$\text{C}\text{H}_2$ molecule}
For $\text{C}\text{H}_2$ molecule, the results are shown in Fig.~\ref{CH2}. In this case, the QAE performs better than the VQE. However, the QAE has the similar issue of degeneracy and level crossing at large molecular distances $d$ between C and H atoms,  in particular, greater than $d\gtrsim 2.1\AA$. 
In this calculation, 10 qubits are needed to represent the Hamiltonian.
 We note that one of the earlier results simulating the molecular energy of CH$_2$ via a quantum algorithm was presented in Ref.~\cite{veis2010quantum} using an iterative quantum phase estimation.
 
\subsubsection{$\text{H}_2\text{O}$ molecule}
For $\text{H}_2$O molecule, the QAE approach in the  approximation of Eq.~(\ref{eq:Adiabatic}) already brings reasonably accurate results for the ground state and its energy compared to those of the VQE, as shown in Fig.~\ref{H2O}. However, there is  also the issue of degeneracy around $d\approx 1.76\AA$.  Despite this deviation, the ground-state energy versus the angle $\theta$ at the equilibrium distance by the QAE is very accurate.   In this calculation, 10 qubits are needed to represent the Hamiltonian. We note that  the simulation of an iterative quantum phase estimation for the ground-state energy of H$_2$O was reported in Ref.~\cite{aspuru2005simulated}. The result for  the first singlet excited state was presented in Ref.~\cite{wang2008quantum}. 
A more recent equation-of-motion method building on the VQE ground state allowed access to a few low lying excited states~\cite{ollitrault2019quantum}.

\subsection{Degeneracy and energy level crossing}
\label{sec:degeneracy}
Degeneracy and energy crossings are two main factors where QAE may fail. If the initial Hamiltonian $H_i$ contains  degenerate ground states, and we arbitrarily choose one of the states as the initial state, the final state  can be a superposition of states and may not necessarily be an eigenstate of the final Hamiltonian. If there exists an energy crossing during the evolution, the state may evolve to an excited state rather than stay in the ground state. These two cases can occur  for real molecules.  In our simulations,  energy level crossings in low lying levels only occur when the molecular distances are large, for example, in Fig.~\ref{gap1}a. But for smaller molecular distances, including near the equilibrium position, no such crossings occur, as illustrated in Fig.~\ref{gap}, and the results from the QAE match the exact results very well, except for the CH$_2$ molecule which has 
two levels that are close (and cross at some distance).

We initially suspected that such issues might be improved by adding in Eq.~\eqref{eq:H} another term,  which is of the form $H_X\equiv\sum_{q} \sigma_q^x$  with a strength $\alpha$, as these Pauli X terms do not commute with Pauli Z terms (occurring in $H_i$),
\begin{equation}
\label{eq:Halpha}
H_\alpha(t)=\Big(1-\frac{t}{T}\Big)H_i+\frac{t}{T}H_p+\alpha\Big(1-\frac{t}{T}\Big)\frac{t}{T}H_X,
\end{equation}
where the summation in $H_X$ is over all qubits labeled by $q$'s and $\alpha$ is an adjustable factor. The $H_X$ term does not change the evolution in the beginning and at the end, but may break up the degeneracy and eliminate some energy crossings in the middle of the evolution; see, e.g., Fig.~\ref{gap1}. However, \textcolor{black}{infinitesimal gaps may exist when the degeneracy is broken}, and thus this modification cannot necessarily ensure the adiabatic evolution to find the final ground state. In our simulations, we do see some minor improvement via the QAE method using small $\alpha$ (e.g. 0.1), but the results become worse for large $\alpha$ (e.g. 0.5).  We comment that it is partly due to this reason that  we will study a spectral projection method below and  will demonstrate that the augmented Hamiltonian~(\ref{eq:Halpha}) used in this new approach can yield much improved outcomes.
	
	\begin{figure}[t!]
	\centering
	\flushleft (a)\\ \includegraphics[width=0.45\textwidth]{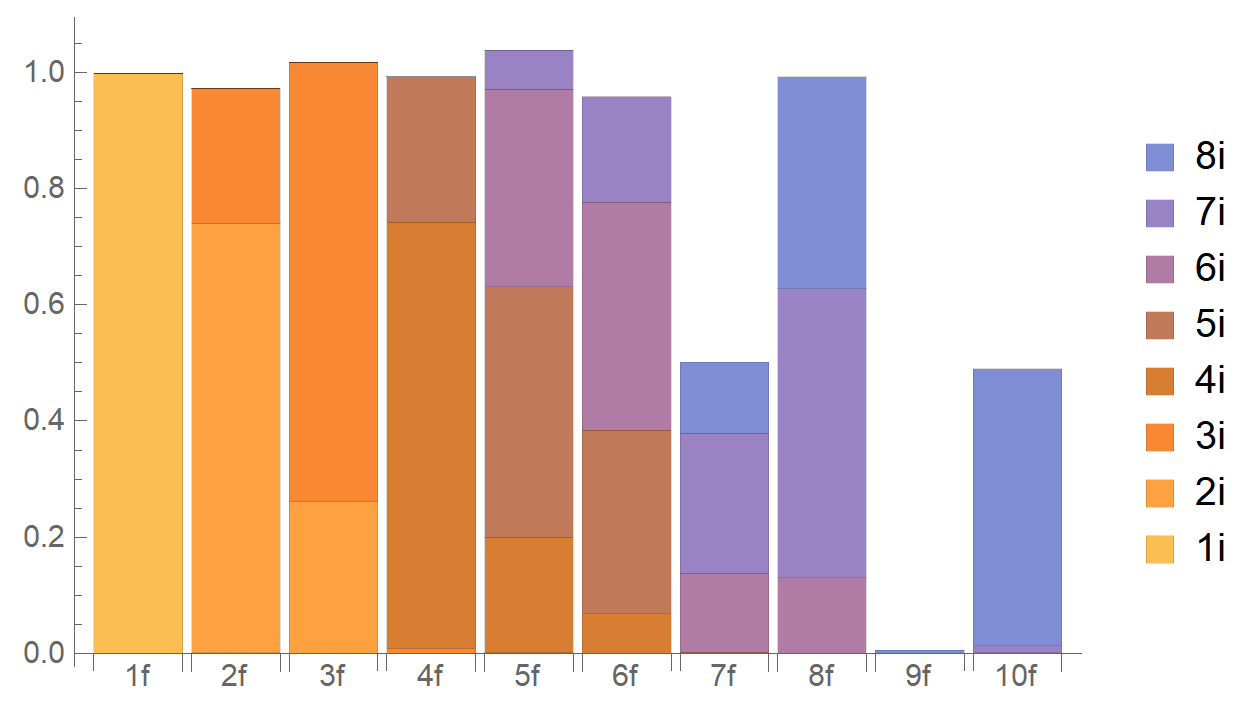}\\
\flushleft	(b)\\ \includegraphics[width=0.45\textwidth]{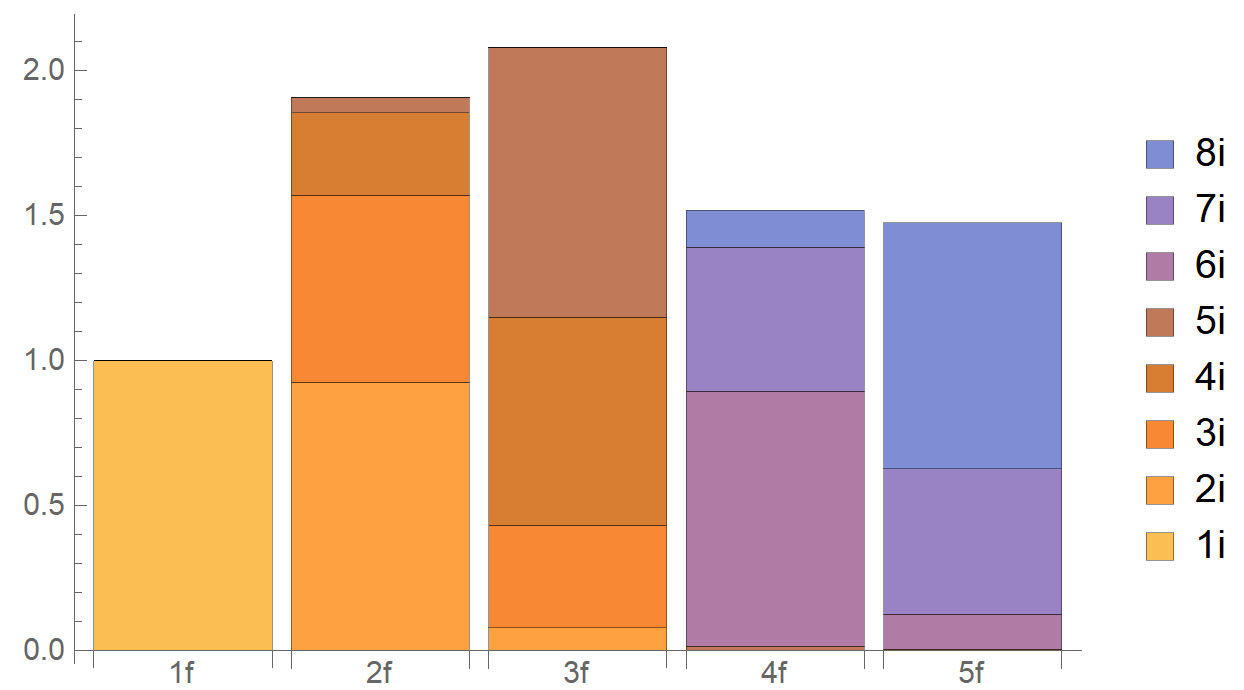}
	\caption{The statistics of how each initial eigenstate will result in after 20 steps of successive eigenstate projections via the QZP method with $\alpha=0$ for (a) $\text{H}_2$O at the equilibrium position $d=0.958 \AA{}$, (b) $\text{BeH}_2$ at  the equilibrium position $d=1.33 \AA{}$. Different colors and their labels 1i, 2i,...8i denote the 8 lowest initial eigenstates. The labels in the horizontal axis 1f, 2f,..., etc.  denote the obtained eigenstates according to the final Hamiltonian and the vertical axis shows the  accumulated distribution (summing those probabilities at a particular final state). These statistics  are gathered from simulating the projection process 1000 times.}
	\label{EPeq}
\end{figure}

\section{Spectral projection method for ground and excited states}
\label{sec:Projection}

\begin{figure*}[ht!]
	\centering
	\includegraphics[width=0.98\textwidth]{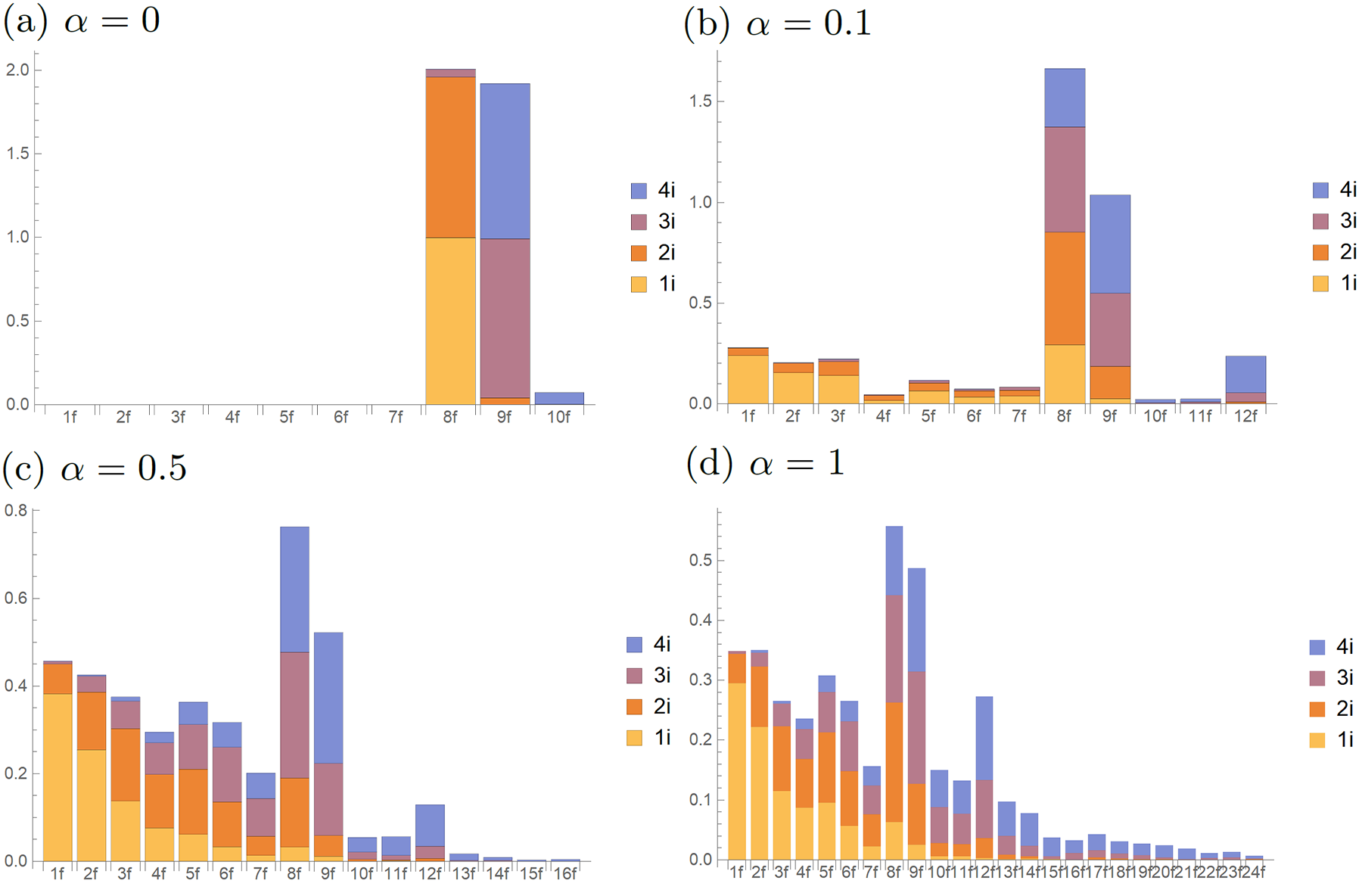}
	\caption{The statistics of which final ground state each initial eigenstate will result in after 20 steps of successive eigenstate projections for  $\text{H}_2$O at the position $d=1.958 \AA{}$  with (a) $\alpha=0$, (b) $\alpha=0.1$, (c) $\alpha=0.5$ and (d) $\alpha=1$. Different colors  and the corresponding labels 1i, 2i,3i, and 4i denote the lowest four initial eigenstates. The labels in the horizontal axis 1f, 2f,..., etc. denote the obtained eigenstates of the final Hamiltonian and the vertical axis shows the accumulated distribution (summing those probabilities at a particular final state from different initial states). Statistics are obtained from simulating the procedure 1000 times. }
	\label{EP}
\end{figure*}
From previous discussions, we see that the QAE method does not always yield accurate results for ground states at large molecular distances due to the limitation of the adiabatic approach that we explain above. For excited states, energy crossings  are more likely to appear during the evolution and the QAE becomes insufficient in reaching low lying excited states of the final Hamiltonian. Therefore, we propose a different approach by using measurement instead of evolution according to the path-dependent Hamiltonian~(\ref{eq:Halpha}). This method was introduced in Ref.~\cite{somma2008quantum}, and termed the quantum simulated annealing in the context of optimizing a classical function. The standard quantum phase estimation~\cite{nielsen2002quantum} can be used to achieve the measurement that projects to the energy eigenbasis, but it is not yet suitable for current noisy quantum computers. Other ways of spectral projection have been proposed, including a quantum-walk based algorithm~\cite{poulin2009preparing} and a Hadamard-test spectral projection method by measuring an ancilla iteratively~\cite{chen2020quantum}.  These different methods also allow extraction of the corresponding eigenenergies. Such  a Zeno-like measurement projects an arbitrary initial state to an eigenstate according to the Born rule. In our work here, we do not specify which particular algorithm will realize this projection, but assume it can be performed.

We discretize the Hamiltonian~(\ref{eq:Halpha}) as follows,
\begin{equation}
	H_k=H_\alpha\left(\frac{k}{N}T\right), \ \mbox{with}\,\, k=0,1,...,N,
\end{equation}
and perform successively the projections in the energy eigenbasis of these Hamiltonians $H_k$. If the overlap of successive ground states is sufficiently close to unity, then by the quantum Zeno effect the resulting final state, after the whole sequence of projections, will be very close to the ground state of  the final Hamiltonian~\cite{somma2008quantum}. If there does not exist any ground-state degeneracy, the projections will drive the initial ground state into the final ground state of the problem Hamiltonian with high probability. 
In the case of the initial degeneracy,  the projection method can yield either one of the possibly split final eigenstates, in contrast to the adiabatic evolution which produces superposition.  A few repetitions of this QZP procedure can result in  \textcolor{black}{multiple distinct} eigenstates.  To obtain the lowest $k$ eigenstates, we prepare about $k$ different lowest initial eigenstates and perform the QZP procedure multiple times. With a high probability, the outputs will contain the desired low lying eigenstates.

\subsection{QZP procedure and numerical results}
Let us summarize the procedure of our quantum Zeno-like projection method as follows:
\begin{quote}
	1. Given the Hamiltonian of a molecule, transform the  fermion operators to Pauli operators;\\
	2. Find the (approximate) maximal set of the Pauli operators by a greedy algorithm and hence the MC Hamiltonian as well as the time-dependent Hamiltonian;\\
	3. Discretize the time steps 	by $t_k=\frac{k}{N}T$ and obtain a series of Hamiltonians $H_k=H_\alpha(t_k)$ at these time steps;\\
	4. Choose one of the MC Hamiltonian's eigenstates as the initial state, and perform the  projections on $H_k$ successively for $k=1,\dots,N$ to obtain one eigenstate of the final Hamiltonian;\\
	5. Repeat the above procedure multiple times to obtain a distribution of final eigenstates. We can pick the desired eigenstate from the distribution for further analysis.
\end{quote}

\begin{figure}[t]
	\centering
	\includegraphics[width=0.5\textwidth]{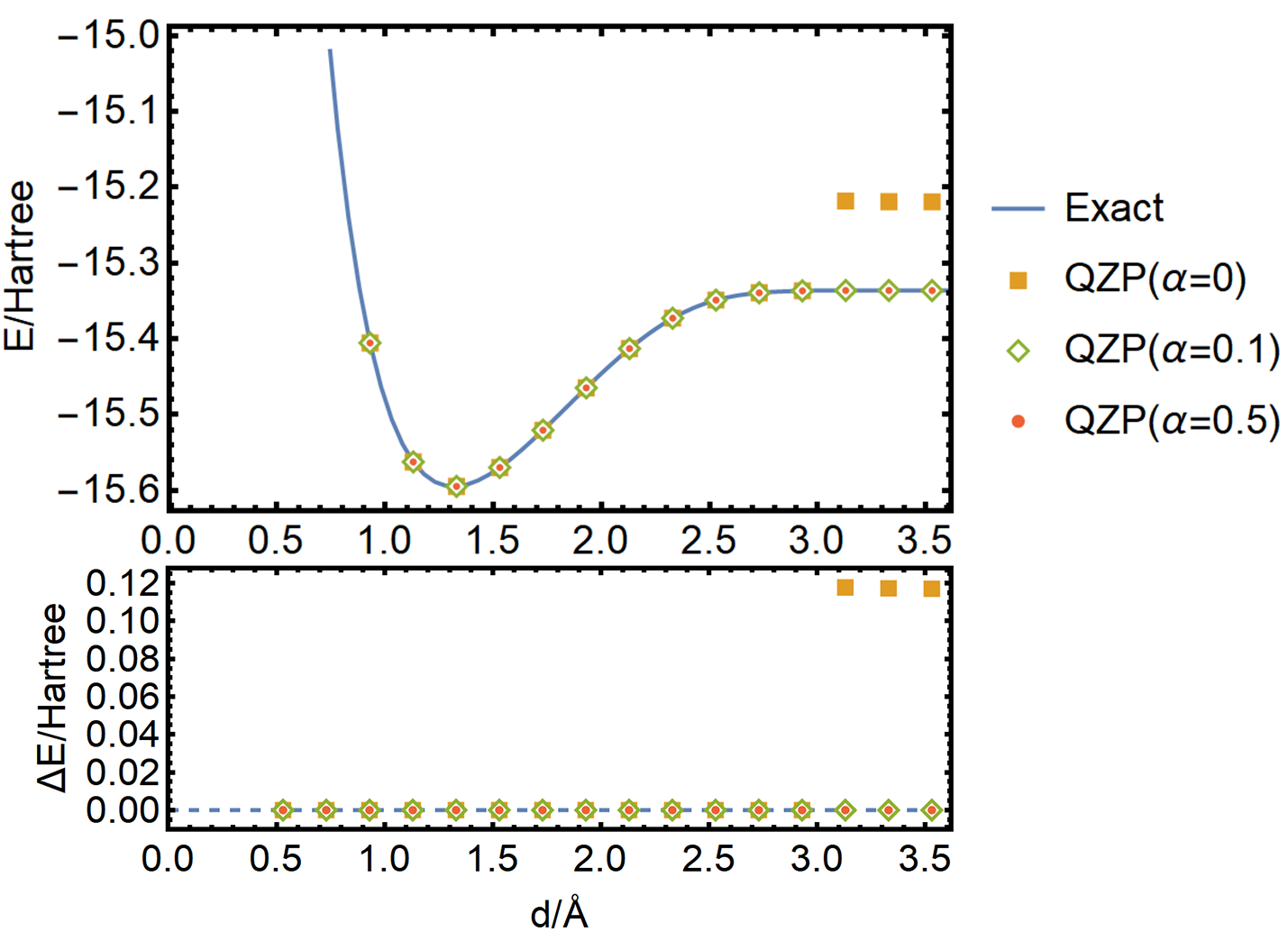}
	\caption{The ground state energy of Be$\text{H}_2$ vs. the distance $d$ between the Be and one H atoms at the equilibrium angle, via the QZP approach with several choices of $\alpha$. Those points on the dashed line are 0-error in terms of machine precision.   
}
	\label{BeH2_alphas}
\end{figure}

 As seen in Fig.~\ref{LiH}, LiH is the simplest of all molecules considered in this paper, and the results from the three different methods all work very well, including the QZP approach discussed in this section. This approach also works well for other molecules, as shown in Figs.~\ref{BeH2},~\ref{CH2}, and~\ref{H2O}, and is the best among the three different methods. 
 This can be understood from the large gap in the path-dependent Hamiltonian, as seen, e.g., in Figs.~\ref{gap} and~\ref{gap1} with $\alpha=0$ and $\alpha=1$. Moreover, we have performed simulations of the QZP method using the lowest few initial eigenstates and gathered statistics of final eigenstates, as illustrated in Fig.~\ref{EPeq} for H$_{2}$O and BeH$_{2}$ at their respective equilibrium position.
 
 \subsection{Improvement using nonzero $\alpha$}
 However, for distances larger than the equilibrium position, the results via QZP with $\alpha=0$ can deviate significantly from their exact counterparts. As explained earlier, this is due to many closely spaced energy levels at large distances that are  degenerate in the infinite separation limit, as illustrated in Fig.~\ref{gap1}. Therefore, the interpolation used in Eq.~(\ref{eq:H}) encounters level crossings and closely packed energies for $s$ close to unity.
\begin{figure}[t]
	\centering
	\includegraphics[width=0.5\textwidth]{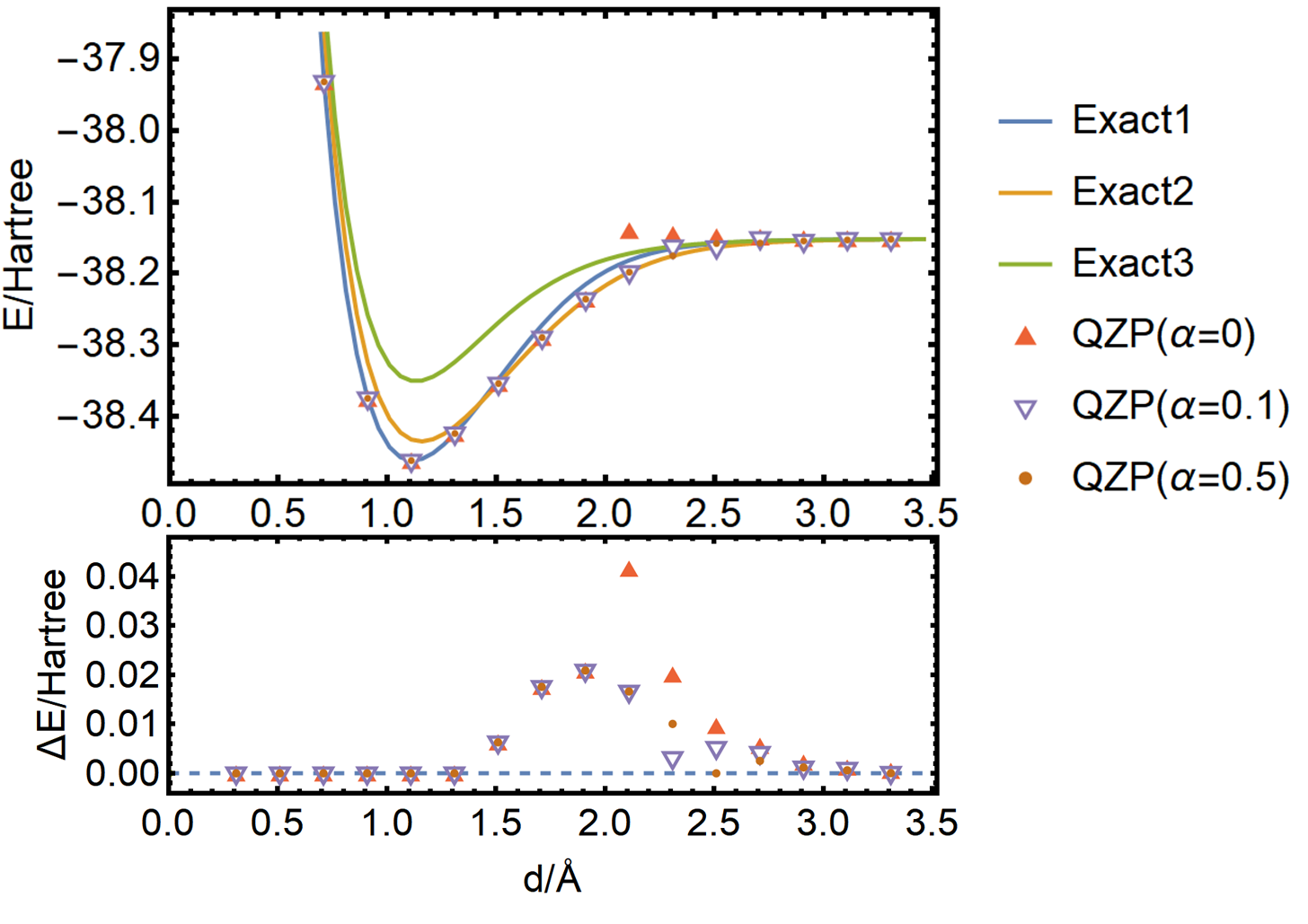}
	\caption{The ground state energy of CH$_2$ vs. the distance $d$ between the C and one H atoms at the equilibrium angle, via the QZP approach with several choices of $\alpha$.   Those points on the dashed line are 0-error in terms of machine precision (note the two points around d=3.0$\AA$ are not on the dashed line).
}
	\label{CH2_alphas}
\end{figure}
As we have indicated, this can be resolved by introducing a potentially degeneracy breaking term, as introduced in Eq.~(\ref{eq:Halpha}), with an overall constant $\alpha$. 

To discuss the effect due to $\alpha$, we thus compare the outcomes using several nonzero $\alpha$'s with those obtained using $\alpha=0$ via the QZP method. As an illustration we compare the statistics of the final states from a few lowest initial states of the MC Hamiltonian with different values of $\alpha$, for the  H$_{2}$O molecule at $d=1.958\AA$, for which the deviation from the exact 
solution is visible in Fig.~\ref{H2O}a. As shown in Fig.~\ref{EP}, the statistics for the lowest few final states increases when $\alpha$ becomes nonzero and the probability of obtaining the final ground state by using the initial ground state of the MC Hamiltonian is enhanced.

\begin{figure}[t]
	\centering
	\includegraphics[width=0.5\textwidth]{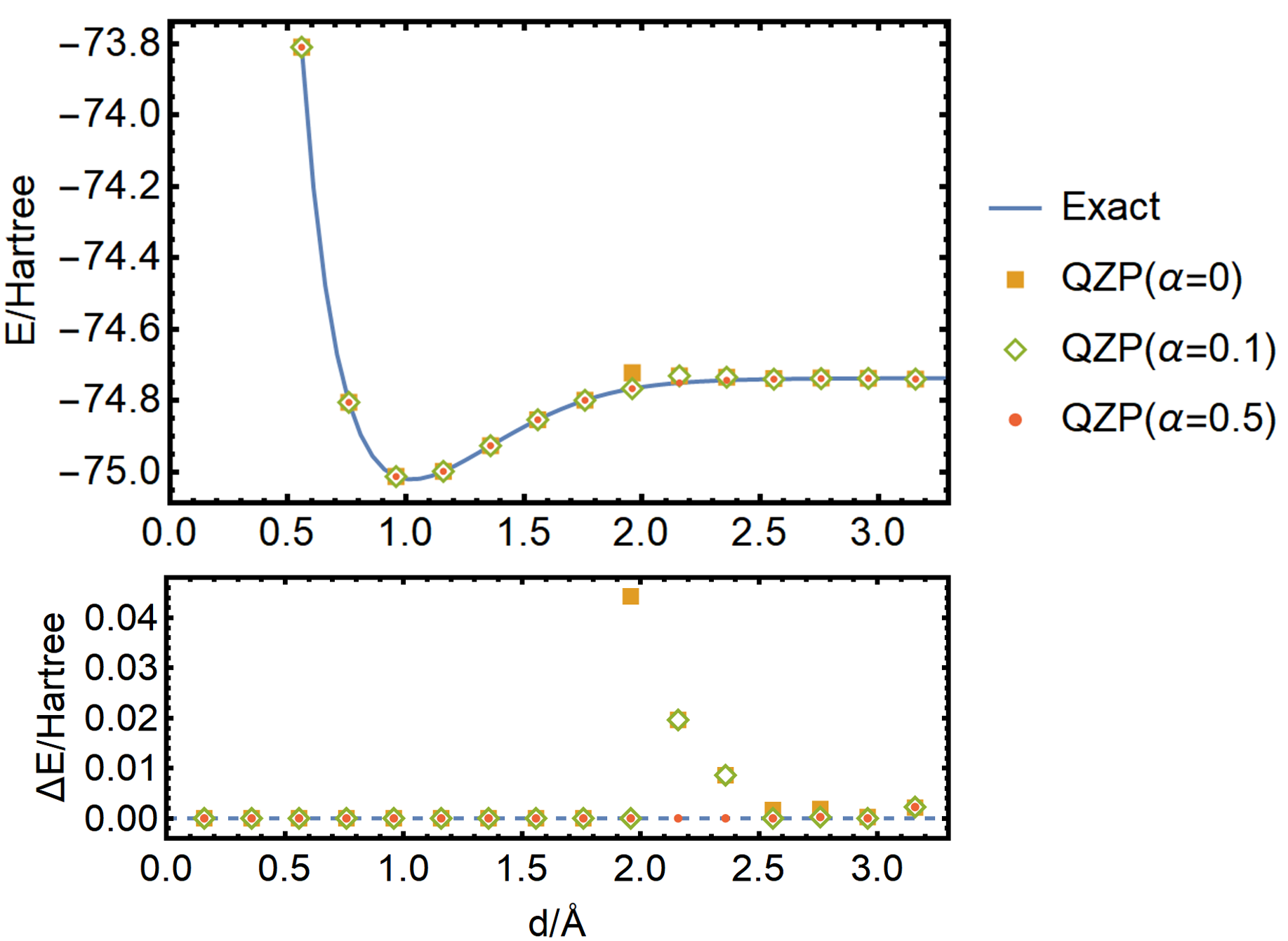}
	\caption{The ground state energy of $\text{H}_2$O vs. the distance $d$ between the O and one H atoms at the equilibrium angle, via the QZP approach with several choices of $\alpha$. Those points on the dashed line are 0-error in terms of machine precision.
}
	\label{H2O_alphas}
\end{figure}

We thus see improvements of the QZP method in obtaining the ground states, as shown in Figs.~\ref{BeH2_alphas},~\ref{CH2_alphas}, and~\ref{H2O_alphas}. There, we only use the initial ground state as the input and repeat the procedure for 40 times  selecting the lowest energy. By contrast, the QAE cannot be improved by such repetition; moreover, the $\alpha$ term may even make the results of QAE method worse (not shown). For the QZP method, the choice of the exact value of $\alpha$ does not seem to be important for the overall performance. If necessary, one may repeat the QZP procedure with a few different values of $\alpha$. Thus, we conclude that the QZP method with nonzero $\alpha$ provides an alternative and viable  approach for studying molecular energies.

\begin{figure}[t]
	\centering
	\includegraphics[width=0.5\textwidth]{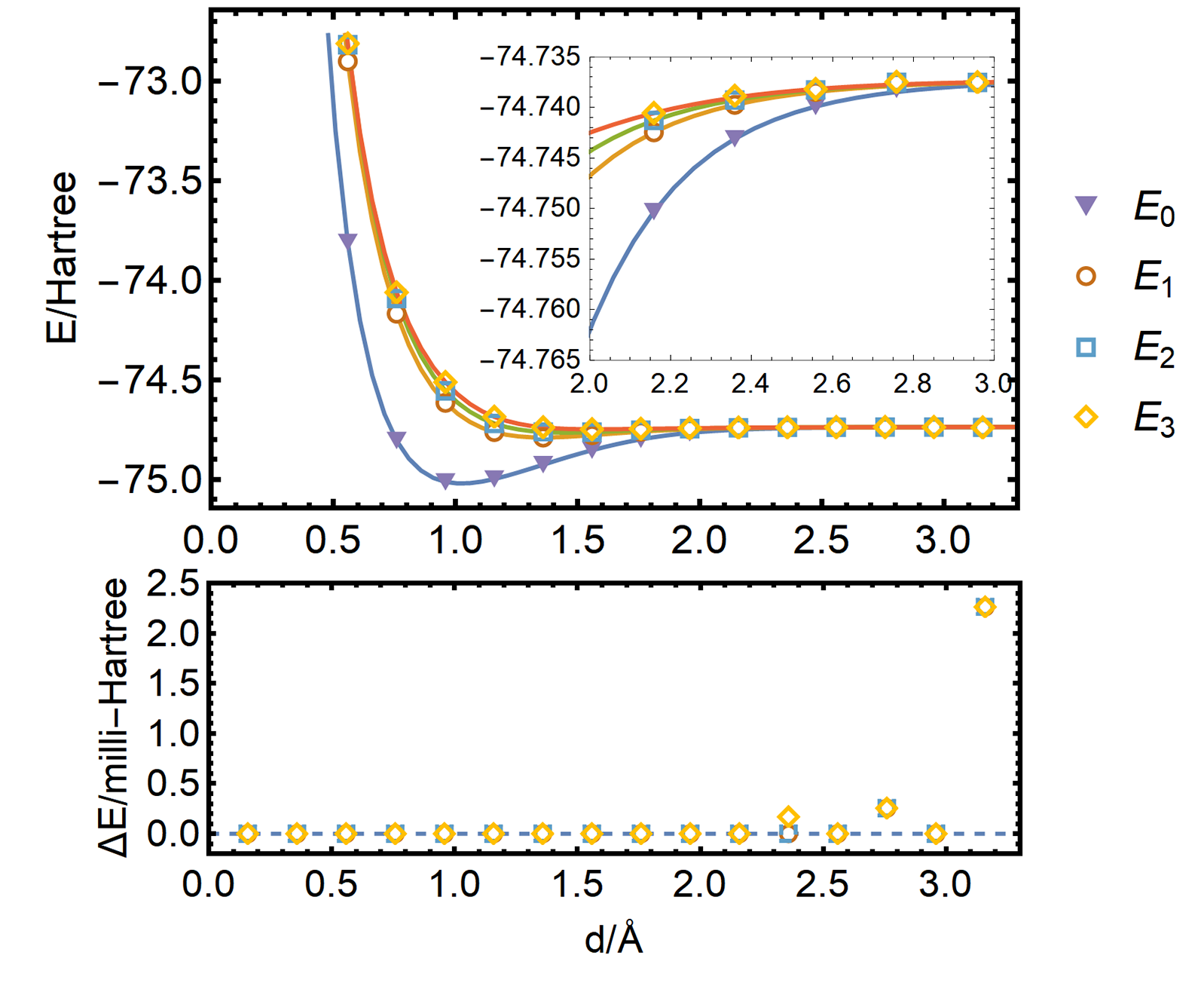}
	\caption{The energy of the ground state and first three excited states for $\text{H}_2$O calculated by the QZP with $\alpha=0.5$. The inset shows the blow-up in the range $d\in[2.0,3.0]\AA$. The lines indicate the exact solution.  We use the lowest 4 eigenstates of MC Hamiltonian as initial states, and repeat the projection procedure 40 times  taking the lowest 4 finial eigenstates as results. Those points on the dashed line are 0-error in terms of machine precision.}
	\label{Ee}
\end{figure}

 We also demonstrate the utility of the QZP method for the lowest few eigenstates, using the H$_2$O molecule as an illustration.  In order to do this, we choose as initial states the lowest few eigenstates of the MC Hamiltonian~(\ref{eq:MSH}). Repeating the procedure, we obtain  and present the lowest four energy values \textcolor{black}{as} a function of the O-H distance, shown  in Fig.~\ref{Ee}.  In this calculation, we use $\alpha=0.5$ in the augmented Hamiltonian. The results are accurate as seen from the lower panel in Fig.~\ref{Ee}, where the errors are on the order of milli-Hartree or even smaller.

While the QAE method in Sec.~\ref{sec:AQC} may result in superposition of final eigenstates, the projection method by design will always result in an eigenstate, despite the fact that we cannot predict in advance which eigenstate will appear. If the path-dependent Hamiltonian always possesses a gap between $s=0$ and $s=1$, then the Zeno effect drags the initial ground state to the final ground state. The introduction of the $\alpha$ term in the Hamiltonian reduces the energy level crossing (e.g. see Fig.~\ref{gap1}b) and improves the performance of the QZP method. 
Regarding the excited states via the QZP method, we note that their fidelity does not depend on the fidelity of the ground state, in contrast to the VQE approach with the equation of motion~\cite{ollitrault2019quantum}.

\begin{figure}[t]
	%\centering
	\flushleft
	(a)\\
	\hspace{0.5cm}	\includegraphics[width=0.38\textwidth]{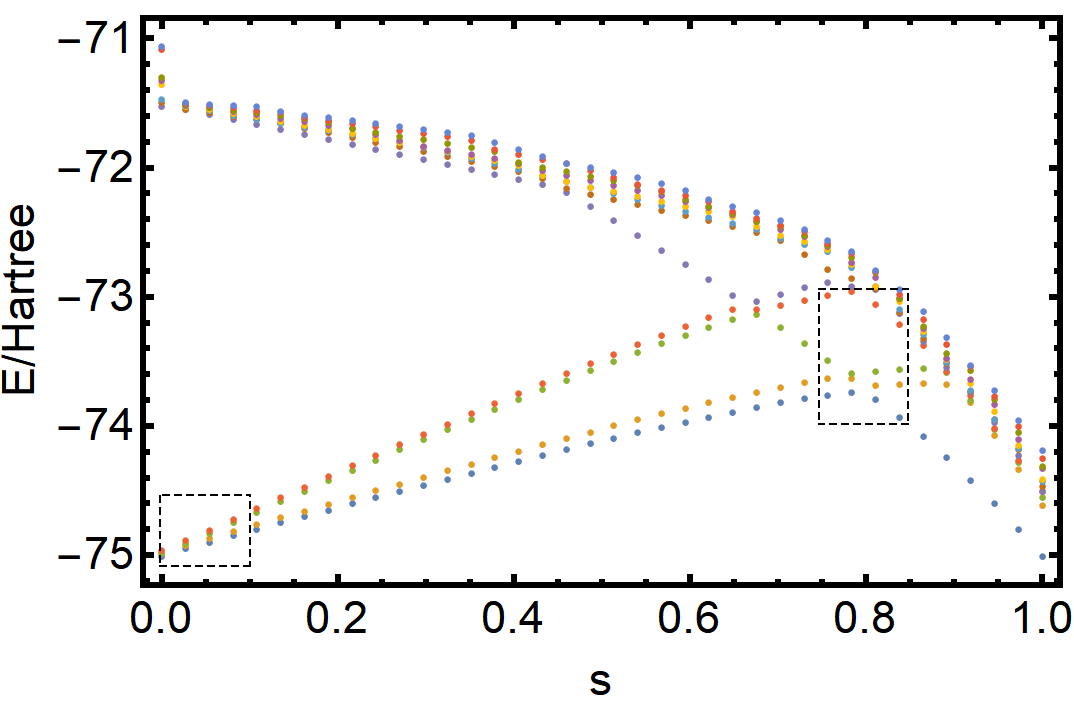}\\
	\flushleft (b)\\
		\includegraphics[width=0.426\textwidth]{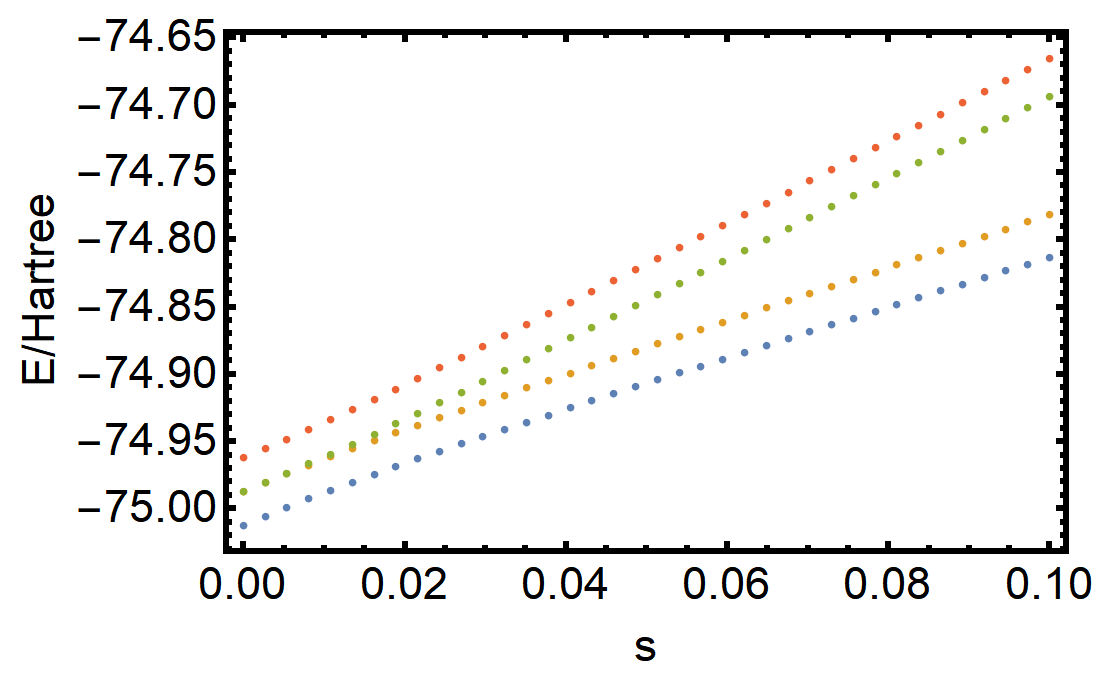}\\
	\flushleft	(c) \\
	\hspace{0.2cm}	\includegraphics[width=0.4\textwidth]{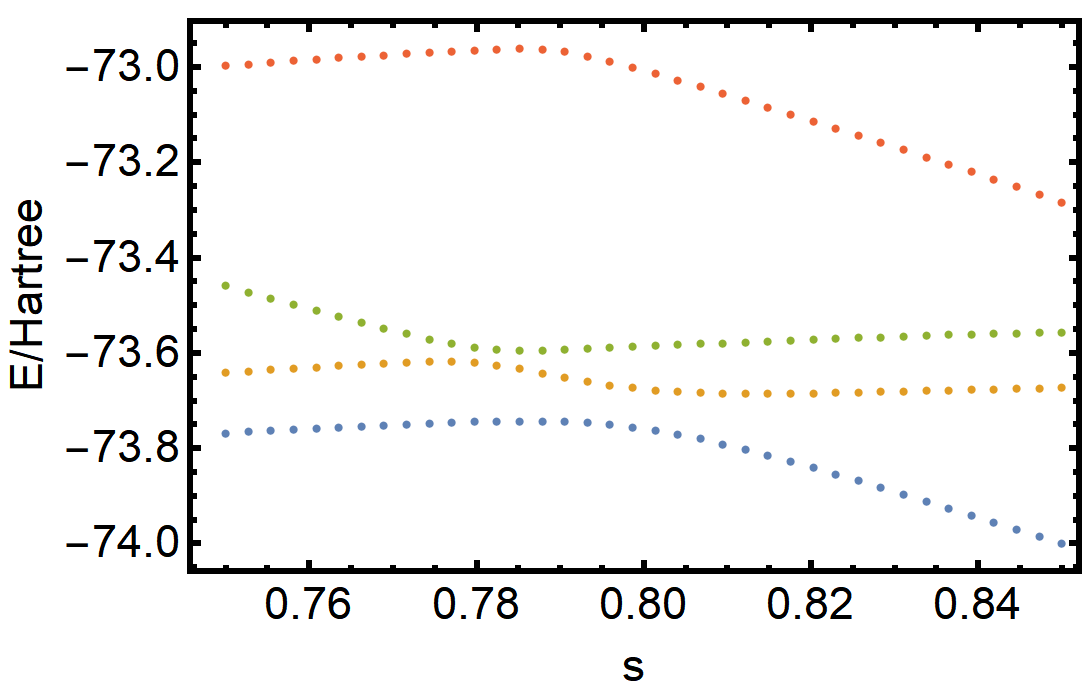}
	\caption{The  energy of the lowest 12 eigenstate of $H(s)=(1-\frac{t}{T})H_{HF}+\frac{t}{T}H_p$, for $\text{H}_2$O at the equilibrium position. Panel (a) shows the whole range of $s\in[0,1.0]$, whereas panels (b) and (c) show the blow-up in some regions of $s$.}
	\label{HF}
\end{figure}
\section{Comparison with Hartree-Fock initial Hamiltonian}
\label{sec:HF}

In the previous sections, we use the maximum commuting Hamiltonian as the initial Hamiltonian and its ground state as the initial state. From the perspective of the commonly used Hartree-Fock approximation, it may seem more natural to use the Hartree-Fock approximated ground state as the initial state and its associated Hamiltonian (such that the former is the ground state of the latter), i.e., the qubit-version of the Fock operator~\cite{mcardle2020quantum}, as the initial Hamiltonian $H_{HF}$ in constructing the full time-dependent Hamiltonian $H(t)$. In doing this, we convert the fermionic operators to qubit operators.

 We thus numerically calculated the energy levels of $H(s)=(1-\frac{t}{T})H_{HF}+\frac{t}{T}H_p$ for $\text{H}_2$O at the equilibrium position, see Fig.~\ref{HF}. The Fock operator $H_{HF}$  is  computed by the {\tt PySCF} package~\cite{PYSCF} and transformed to Pauli operators by the {\tt Qiskit} package~\cite{Qiskit}. The transformation includes parity mapping and the freezing of the $1s$ orbital. For visual convenience we add an identity term to this  Hartree-Fock Hamiltonian to force its ground state energy to be equal to the ground state energy of the final Hamiltonian. We find that the spectra along such  \textcolor{black}{an interpolation is not favorable in comparison to the one with the MC Hamiltonian as the initial Hamiltonian}, as there are low-lying eigenstates with close energies and small gaps or level crossings, as well as degeneracies for a certain range of $s$.  Adding the $\alpha$ term would not improve upon matters here. This shows that using the MC Hamiltonian as the initial Hamiltonian will work better than the Fock operator. 

\section{Concluding remarks}
\label{sec:conclude}
We propose to use an adiabatic framework {to compute}  eigenstates and energies of molecules, and, in particular, we employ the maximal commuting set to construct the initial Hamiltonian, as opposed to, e.g., the Hartree-Fock Hamiltonian.  
However, the issue of degeneracy can cause instabilities in the adiabatic quantum-computational approach (i.e., the QAE), where the ground-state energies of molecules cannot be calculated accurately in the event of the bond breaking at large molecular distances. As we have examined, the alternative projection method, i.e., the quantum Zeno-like projection (QZP) approach with additional Pauli X terms, resolves this issue and obtains very accurate results.   

We remark that it is also possible to apply the projection method in between some  steps of  the adiabatic evolution or even at the end. However, it is not clear a priori where to insert such projection steps. We have  checked whether the projection added to the end of the QAE method could improve our results. We found that in the case of bond breaking, such as that considered in Fig.~\ref{EP}, this additional projection does not help to arrive at the ground state or other low lying states, even if the augmented Hamiltonian $H_\alpha$ in Eq.~(\ref{eq:Halpha}) is used. Namely, the distribution of the final eigenstates has little probability in the final  ground state and lowest few eigenstates. The QAE method followed by an eigenstate projection is thus not equivalent to the QZP method. This result also illustrates that although the quantum simulated annealing is generally equivalent to quantum adiabatic computation, the use of the former can give rise to an advantage in comparison with the latter in the application of molecular energy.

 We have  demonstrated numerically that the QZP approach can output  low lying eigenstates with high probability in addition to the ground state. Compared to finding excited states via the VQE~\cite{ollitrault2019quantum}, our method does not rely on a precise ground state.  Throughout this work, we have also compared the results from our methods to those of VQE for the ground-state energy,   because VQE has become the standard quantum procedure for molecular energies and reference results can be found in the literature. However, one needs to treat this comparison with a grain of salt, as the VQE is a near-term method and our methods may require longer-term devices. 
 
In Ref.~\cite{xia2017electronic}, Xia, Bian and Sais proposed a method to map a quantum chemistry Hamiltonian to an Ising spin-class Hamiltonian at the expense of more qubits with a constant prefactor. In this way, quantum chemistry problems can be carried out on the existing quantum annealers of D-Wave and some experiments have been already carried out~\cite{streif2019solving}. Current quantum annealers have constraints on connectivity and types of implementable coupling and thus our QAE approach is not yet directly implementable on these machines. However, one may consider digital implementations of adiabatic quantum computation, akin to our simulations of the QAE approach. Given the number $n_{\rm ob}$ of electronic orbitals used, there are ${\cal O}(n_{\rm ob}^4)$ interaction terms in the Hamiltonian. To digitize the evolution $e^{-i H(t)\Delta T}$, one has to use approximately at most ${\cal O}(n_{\rm ob}^4)$ terms  in the first-order Trotter decomposition; this number can be reduced by dividing these Hamiltonian terms into groups, where within each group, \textcolor{black}{all elements commute; see e.g~\cite{1907.13623}.} There is also an overhead to turn multi-qubit gates into some combination of one- and two-qubit gates. To complete the digital implementation of the adiabatic evolution, one needs to repeat such gate sequences at $T/\Delta T$ time steps (which is 20 in most of our simulations).   
In current digital quantum computers, such a long sequence of gate \textcolor{black}{operations} may suffer substantially from decoherence and noise. Encouragingly,
 Ref.~\cite{barends2016digitized} reports an experimental realization of the adiabatic method with an execution of over 1000 gates, and where experiment could successfully find the solution to random instances of the one-dimensional Ising problem. This study indicates that if the gate error rates can be further improved, the adiabatic approach could be promising for complex quantum chemistry problems on near-future large-scale quantum devices.
 
 At this moment, our QZP approach is also not readily implementable due to the requirement to project onto the energy eigenbasis, and the Hamiltonian involves  multiple-qubit Pauli terms.  Physical systems that likely possess multiple-qubit gates include  trapped ions~\cite{shapira2020theory} and Rydberg atoms using blockades~\cite{young2020asymmetric}. But projection of these qubits onto eigenstates of a multi-qubit interacting Hamiltonian is still challenging. The standard phase estimation algorithm~\cite{nielsen2002quantum} can achieve this but it needs to use controlled evolution of the form $c-U^{2^k}$, \textcolor{black}{ where $U=e^{-iH\Delta T}$ is the evolution operator, `$c\, -$' denotes the controlled operation, and $k=0,\cdots n_a-1$ where $n_a$ is the number of ancillary qubits and sets the accuracy in  the energy and  the projection.} Such a controlled evolution is the bottleneck of the algorithm as the subsequent inverse quantum Fourier transform can be implemented semi-classically~\cite{griffiths1996semiclassical}.  Similar to the above analysis in the QAE approach, the implementation of $c-U^{2^k}$ requires decomposing at most ${\cal O}(n_{\rm ob}^4)$ terms to single- and two-qubit gates. It incurs slightly more overhead as the controlled version involves one more ancillary qubit for each term.  However, the approximation in the Trotter decomposition destroys the perfect exponential relation $2^k$ in the power of $U$ and the desired outcome of ancillary qubits registering a phase in the form $|0\rangle + e^{i \phi 2^k} |1\rangle$.  An iterative version of the phase estimation that uses a single ancilla may be used~\cite{dobvsivcek2007arbitrary}, but it also suffers from the same drawback if the exponentiation of $U$ is not exact. 
 
 On the other hand, there are a few other proposals to perform spectral projection~\cite{somma2008quantum,poulin2009preparing,chen2020quantum}, and it is an interesting future direction to consider how to make spectral projections for Hamiltonians with multiqubit Pauli terms suitable for near-term quantum computers. For example, one may use the specific proposal of the Hadamard-test-like method in Ref.~\cite{chen2020quantum}. There, the ancilla's initial state does not need to be $|+\rangle$ but can be an arbitrary pure state, and the evolution time $\Delta T$ \textcolor{black}{need} not be proportional to $2^k$ and can even be random. It was demonstrated by numerical simulations  that Trotterization did not affect the projection.  Thus, our QZP approach with the spectral projection implemented in this fashion  has  similar time complexity as the iterative phase estimation approach but it has the advantage of being robust to  errors in both timing and Trotterization. To reduce the use of multi-qubit gates, one can thus optimize the expansion of the controlled evolution into discrete one- and two-qubit quantum gates. \textcolor{black}{The number of gates} depends on the native gates in particular physical systems and requires further investigation.  
 With future larger-scale quantum computers and better multi-qubit gates~\cite{young2020asymmetric}, the Zeno-based approach may be a viable approach for quantum chemistry of large molecules.

\begin{acknowledgments} This work was supported by
the National Science Foundation under grant No. PHY 1915165 and an SBU-BNL seed grant.  We acknowledge the use of the Qiskit package~\cite{Qiskit}  in obtaining the molecular Hamiltonians in the qubit form and in simulating the VQE procedure.  We thank   Gabriel Kotliar and Qin Wu for useful discussions and  Robert Konik and Deyu Lu for providing valuable comments on our manuscript. 
\end{acknowledgments}
\bibliography{Mstab}{}
\end{document}